\pdfoutput=1 
\documentclass{JINST}
\usepackage{xspace}
\usepackage{multirow}

\def\vx {\mbox{\ensuremath{\vec{x}}}\xspace}

\title{Machine learning and multivariate goodness of fit} 

\author{Constantin Weisser and Mike Williams\\
Laboratory for Nuclear Science, Massachusetts Institute of Technology, Cambridge, MA 02139}

\abstract{
 Multivariate goodness-of-fit and two-sample tests are important components of many nuclear and particle physics analyses.
While a variety of powerful methods are available if the dimensionality of the feature space is small, such tests rapidly lose power as the dimensionality increases and the data inevitably become sparse. 
Machine learning classifiers are powerful tools capable of reducing highly multivariate problems into univariate ones, on which commonly used tests such as $\chi^2$ or Kolmogorov-Smirnov may be applied. 
We explore applying both traditional and machine-learning-based tests to several example problems, and study how the power of each approach depends on the dimensionality.
A pedagogical discussion is provided on which types of problems are best suited to using traditional versus machine-learning-based tests, and on the how to properly employ the machine-learning-based approach.  
}

\begin{document}

\section{Introduction}

 Multivariate goodness-of-fit and two-sample tests are important components of many nuclear and particle physics analyses.
For example, an unbinned maximum likelihood fit of a probability density function (PDF) to a data sample is commonly performed in such analyses.  
The maximum likelihood value cannot be used to determine the level of agreement between the fit PDF and the data; therefore, some other goodness-of-fit test must be employed for this important task.
A related problem is determining whether two data samples --- of which, one may be simulated --- share the same parent PDF, where the PDF itself may be unknown, {\em i.e.}\ no parametric expression of the PDF is required. 
The most widely used statistic in nuclear and particle physics analyses for both goodness-of-fit and two-sample testing is the $\chi^2$\,\cite{Pearson1900}, though many other tests are also applicable to both univariate and multivariate problems (see, {\em e.g.},  Refs.~\cite{GOF,Williams:2010vh}). 
While a variety of powerful tests exist when the dimensionality of the feature space is small, all such tests rapidly lose power as the dimensionality increases and the data inevitably become sparse; this is known as the {\em curse of dimensionality}~\cite{Bellman}. 

Machine learning (ML) classifiers are powerful tools capable of reducing highly multivariate problems into univariate ones, on which commonly used tests such as $\chi^2$ or Kolmogorov-Smirnov (KS)~\cite{tKOL33a,Smirnov} may be applied.
To the best of our knowledge, this idea was first proposed in the literature in Ref.~\cite{Friedman:2003id}. 
It has since become a popular method in the data-science toolkit\,(see, {\em e.g.}, Ref.~\cite{Lopez-Paz2016} and references therein, especially Refs.~\cite{Gretton2012,Ramdas2016}) for the following reasons: 
ML classifiers are well suited to problems where the dimensionality is large; 
powerful ML classifier algorithms are now available in easy-to-use open-source software packages, and in several programming languages;
the previous two points have made the usage of ML methods commonplace in classification problems, which in turn has led to analysts developing both skill and comfort using them; 
and, finally, once the ML classifier has reduced the problem into a univariate one, quantifying the level of agreement with a $p$-value is simple, {\em e.g.}, using the well known $\chi^2$ or KS test. 

Despite the fact that ML algorithms are now commonly used in nuclear and particle physics analyses, {\em e.g.}, to classify events as signal or background, including in online data-selection {\em trigger} algorithms~\cite{Gligorov:2012qt}, most of the community appears to be unaware of the possibility of employing ML classifiers for goodness-of-fit or two-sample testing.
In this article,  we explore applying both traditional and ML-based methods to various example problems, and study how the power of each depends on dimensionality.
A pedagogical discussion is provided on which types of problems are best suited to the use of traditional versus ML-based approaches to goodness-of-fit and two-sample testing, and on how to properly employ the machine-learning-based approach. 
The primary goals of this article are: 
\begin{itemize}
\item to present the ML-based approach to the nuclear and particle physics community in a clear and concise manner, using language familiar to members of our community;
\item to provide a step-by-step guide to proper application of the ML-based approach, including the inclusion of systematic uncertainties;
\item and
to impart in the reader some intuition of when to expect ML-based goodness-of-fit tests to outperform traditional methods.
\end{itemize}
This article is structured as follows:
Sec.~\ref{sec:traditional} discusses traditional testing methods;
an overview of applying ML algorithms for these tasks is presented in Sec.~\ref{sec:ml}; 
example problems are presented in Sec.~\ref{sec:ex};
and a summary is provided in Sec.~\ref{sec:sum}.

\section{Traditional Methods}
\label{sec:traditional}

This section provides a brief overview of traditional, {\em i.e.}\ non-ML-based, methods for performing goodness-of-fit and two-sample testing. For a more complete discussion, we refer the interested reader to Refs.~\cite{GOF,Williams:2010vh}. 
Throughout this article, we denote the $d$-dimensional vector of event features as \vx and PDFs as $f(\vx)$, each with subscripts where necessary for clarity. 

\subsection{Goodness-of-Fit \& Two-Sample Testing}

The goal of a goodness-of-fit test is to obtain a $p$-value for the hypothesis $f(\vx)=f_0(\vx)$ given observations $\{\vx_i\}_{i=1}^{n}$ sampled from $f$, where $f_0$ is some test PDF, {\em e.g.}, $f_0$ may have been determined by performing an unbinned maximum likelihood fit to the data. 
The $p$-value is the probability of observing a test statistic that corresponds to lesser agreement than the actual observed value when the null hypothesis $f=f_0$ is true.
Therefore, if $f=f_0$, the $p$-values are uniformly distributed on the interval $(0,1)$. 
A related problem is determining whether two data samples, $\{\vx_i\}_{i=1}^{n_a}$ and $\{\vx_j\}_{j=1}^{n_b}$, share the same parent PDF, {\em i.e.}, the goal is to determine the $p$-value for the hypothesis $f_a = f_b$, where parametric expressions of $f_a$ and $f_b$ may not be available, {\em e.g.}, searching for $CP$ violation in Dalitz decays~\cite{Williams:2011cd}. 
In each case, one ideally wants to use a test that has power against all alternatives in the asymptotic limit $n_{(a,b)}\to\infty$. 
{\em N.b.}, one can recast a goodness-of-fit test into a two-sample test by generating a simulated (Monte Carlo) data sample from the PDF $f_0$, then performing a two-sample test on the simulated and experimental data samples. 
This is a common approach in cases where an analytic expression for the detector response does not exist, and is a valid way of assessing the goodness of fit in all cases.

\subsection{Univariate Methods}

For the univariate case, each observation involves a single feature $\vx_i = x_i$. 
The most commonly used one-dimensional tests in nuclear and particle physics are the $\chi^2$ and KS tests; however, a variety of other powerful tests are available for univariate problems, {\em e.g.}, the Anderson-Darling (AD)\,\cite{tAND52a} and Cramer-von Mises (CvM)\,\cite{Cramer,VonMises} tests.  
We assume that the reader is familiar with both the  $\chi^2$ and KS tests and omit any detailed discussion on them; however, we do feel it is important to note that many algorithms for obtaining $p$-values from a two-sample KS test assume that the underlying distributions are continuous.
The response of some ML classifiers is discrete, so some care needs to be applied here.
If the fraction of repeated $x_i$ values is small, then using an algorithm that assumes $x$ is continuous will give a good estimate of the $p$-value; however, if the number of ties is large, the $p$-value must be obtained using another method, {\em e.g.}, the permutation test~\cite{Permute}.

\subsection{Multivariate Methods}

As with the univariate case, there are many methods available for performing multivariate goodness-of-fit and two-sample tests; however, most are (in some way) based on local-density, and any density-based method loses power rapidly as $d$ increases (even power techniques like the energy test~\cite{Energy}). 
For a detailed review of traditional multivariate tests, see Refs.~\cite{GOF,Williams:2010vh}.
The $\chi^2$ test is applicable in any number of dimensions, but it is important to define a binning scheme that avoids producing too many low-occupancy bins when $d$ is large. 
When comparing the $\chi^2$ and ML-based methods below, we employ {\em adaptive} (probability) binning~\cite{ProbBin}, where the binning scheme is defined using the pooled $\{\vx_i\}_{i=1}^{n_a}+\{\vx_j\}_{j=1}^{n_b}$ data sample such that all bins in both the $a$ and $b$ samples are expected to have equal population if $f_a = f_b$ when $n_{a,b}\to\infty$.
The number of bins is optimized separately for each example problem studied in Sec.~\ref{sec:ex}, and we note that this number is in all cases less than one might naively expect.

\section{Machine Learning Methods}
\label{sec:ml}

ML classifiers are powerful tools capable of reducing highly multivariate problems into univariate ones. 
This section describes the procedure for using a ML algorithm to reduce the problem down to a single dimension, then applying a well-known univariate two-sample test to obtain the $p$-value.  
Inclusion of systematic uncertainties and some details about specific ML algorithms is also provided. 
{\em N.b.}, a related topic is the connection between ML classifiers and likelihood ratio testing, see Ref.~\cite{Cranmer:2015bka} for detailed discussion and Ref.~\cite{carl} for implementation in software. 

\subsection{Strategy}

The basic strategy employed is to use a ML classifier to reduce the $d$-dimensional problem down to a one-dimensional one, then to apply a well-known univariate two-sample test to obtain the $p$-value. Specifically, the steps are:
\begin{itemize}
\item each sample is labeled and split into training, validation, and testing subsamples;
\item from the training samples, the ML algorithm learns a model $m(\vx)$ designed to classify which sample an event belongs to based on the measured values of its \vx features ($m(\vx)$ is often referred to as the score or response);
\item from the validation samples, the ML algorithm hyperparameters are optimized;
\item from the testing samples, two univariate distributions $\{m(\vx_i)\}$ are constructed, one for each label value;
\item and, finally, a univariate two-sample test, {\em e.g.}, the $\chi^2$ or KS test, is performed on the two testing-sample $\{m(\vx_i)\}$ distributions to obtain a $p$-value.
\end{itemize}
To instead perform a goodness-of-fit test, the same procedure is followed but where one of the data samples is simulated by sampling from the test PDF $f_0$.  
This procedure is simple to implement as it involves sequentially performing two tasks that physicists already know how to do: (1) training a ML classifier, and (2) obtaining a $p$-value from a comparison of two one-dimensional distributions. 
 {\em N.b.}, if the sample sizes are limited, $k$-fold cross validation can be used. 

\subsection{Systematic Uncertainties}

A simple method for including systematic uncertainties involves first choosing to bin the one-dimensional $\{m(\vx_i)\}$ distributions and selecting the $\chi^2$ two-sample test to obtain the $p$-value. 
This facilitates adding bin-by-bin systematic uncertainties in the standard way one includes them for any $\chi^2$ test ({\em e.g.}, see Eq.~\ref{eq:chi2} below). 
The best approach for determining the systematic uncertainty in each $\{m(\vx_i)\}$ bin will depend on the specific problem, but will be similar to assigning systematics for any binned measurement. 
Some useful techniques include: 
weighting the events in the testing samples to account for systematic effects, then producing weighted $\{m(\vx_i)\}$ distributions where the shifts in the bin contents relative to the unweighted distributions are used to determine the uncertainty; and
generating alternative data samples that include some systematic effect(s), then repeating the entire procedure, including the training, where again the differences in the $\{m(\vx_i)\}$ distributions are used to determine the uncertainties. 

\subsection{Algorithms Considered}

Since our focus is on goodness-of-fit and two-sample tests, we restrict our study to ML algorithms that are easy for physicists to use and do not require significant CPU resources.\footnote{Our motivation here is pragmatic. If performing a test requires purchasing commercial software, dedicating significant effort to learning how to implement the ML algorithm in software, or if substantial CPU power is needed to adequately train the ML algorithm, we do not believe that such a method would be adopted by a substantial fraction of the physics community for use as a goodness-of-fit or two-sample test. Of course, the reader is free --- even encouraged --- to disagree, and could instead apply the ML-based procedure using a deep-learning neural network library, {\em e.g.}\ {\sc TensorFlow}\cite{tensorflow}, or some other cutting-edge ML algorithm. The procedure presented in this paper is independent of which ML classifier is used.}   
Even with this restriction, a large number of ML classifiers are now available in easy-to-use open-source software packages. In our study, we consider:
\begin{description}
\item [BDT] boosted decision trees using either AdaBoost\cite{Adaboost}, as implemented in {\sc SciKit-Learn}\,\cite{sklearn}, or XGBoost\cite{Chen:2016:XST:2939672.2939785}, as implemented in {\sc rep}\,\cite{rep}, and a tunable number of trees (other hyperparameters are also tuned that are specific to the boosting algorithms); 
\item [ANN] artificial neural networks that are multilayered, as implemented in {\sc Keras}\cite{keras} and {\sc SciKit-Learn}, with zero or one hidden layer and a tunable number of neurons per layer;
\item [SVM] support vector machines using the radial basis function kernel as implemented in {\sc SciKit-Learn}, with tunable $C$ and $\gamma$ parameters\cite{sklearn}. 
\end{description}
All ML algorithms are trained on the training subsamples,
the hyperparameters are tuned on the validation subsamples using the Bayesian optimization framework as implemented in the {\sc Spearmint} package\,\cite{spearmint},  
and the performance is evaluated on the testing subsamples (these are all independent data samples).  
For the sake of readability, 
only the best-performing ML algorithm from each category is shown for each example problem in Sec.~\ref{sec:ex}. 

\section{Example Problems}
\label{sec:ex}

In this section, we consider three example problems. 
The first two are chosen for their pedagogical value, and because each is easy to implement for any dimensionality. 
For these, we study how the performance of the $\chi^2$ and ML-based tests depends on $d$.
The third problem is more realistic, and involves searching for a two-body particle decay.  
For each problem, ensembles of data sets are generated each with a sample size of 10k events.\footnote{We use 5-fold validation, where each ML algorithm is trained on 8k events and tested on the remaining 2k.  This process is repeated 5 times such that all 10k events are utilized in both training and testing, but always in an unbiased way. To reduce the CPU time required for training, the hyperparameters are optimized for each algorithm only once per ensemble, then used for data sets in the ensemble.
This should not need to be done in a real-world analysis, where only a single data sample is analyzed. 
} 
The notation used in this section is that in all cases the two data samples being compared have parent PDFs labeled by $f_a$ and $f_b$. The $a$ data sample will always originate from $f_1$, {\em i.e.}\ $f_a = f_1$. When the null hypothesis is true, $f_b = f_1$; however, when it is not true, then $f_b = f_2$. 
The power of each method is assessed by studying how frequently it is able to reject the null hypothesis ($f_a = f_b$) at 95\% confidence level.

We universally assume a systematic uncertainty of 1\% in all regions of feature space, largely just to demonstrate how to incorporate systematic uncertainties.  
Therefore, the two-sample $\chi^2$, where each data sample has the same number of events, is defined as
\begin{equation}
\label{eq:chi2}
\chi^2 = \sum\limits_{k=1}^{n_{\rm bins}} \frac{(a_k - b_k)^2}{a_k + b_k + (0.01\cdot a_k)^2 + (0.01\cdot b_k)^2}, 
\end{equation}
where $a_k(b_k)$ denotes the yield of sample $a(b)$ in the $k^{\rm th}$ bin.  The systematic uncertainties are implemented by randomly sampling  $a_k(b_k)$ from the Gaussian functions $\mathcal{G}(\mu=n_{a,b}^k,\sigma=0.01\cdot n_{a,b}^k)$, where $n_{a,b}^k$ is the generated event count in bin $k$. This is done both for the multivariate $\chi^2$ tests, and for the univariate $\chi^2$ tests performed on the ML-response distributions (of course, implementing systematic uncertainties by smearing the bin contents is not required in a real-world analysis).

\subsection{Gaussian}

\begin{figure}[t]
  \centering 
  \includegraphics[width=0.49\textwidth]{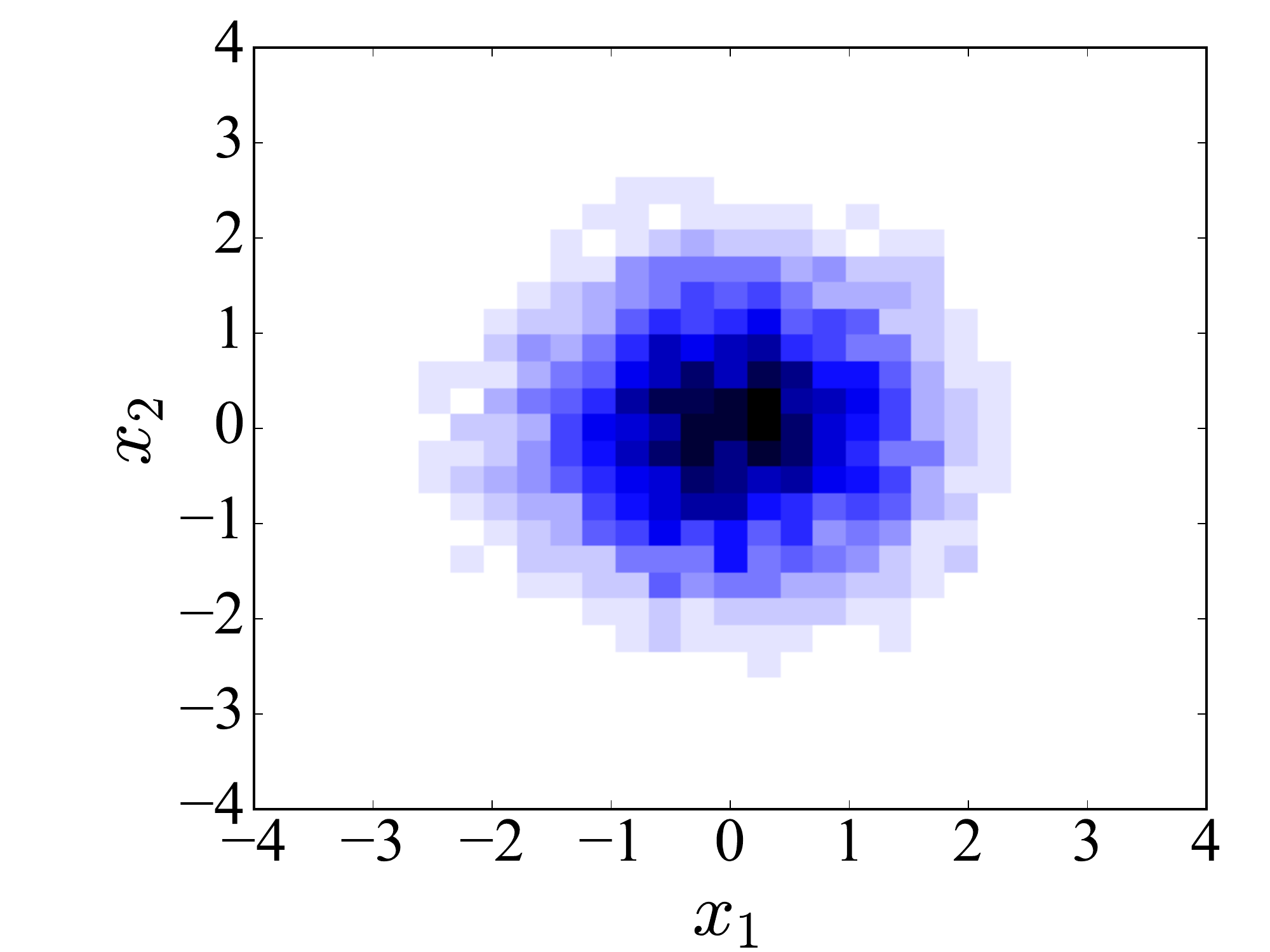}
  \includegraphics[width=0.49\textwidth]{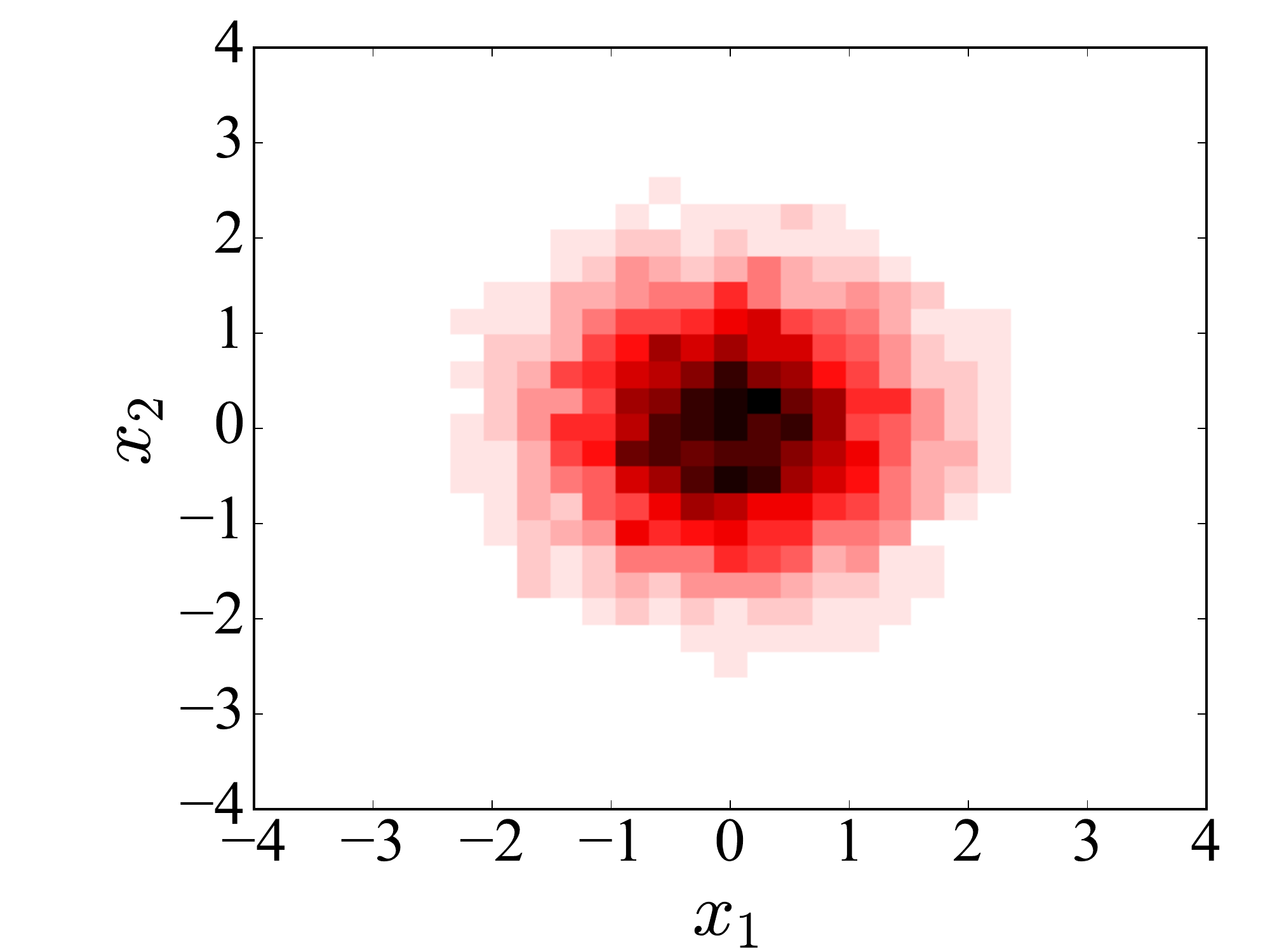}
  \includegraphics[width=0.49\textwidth]{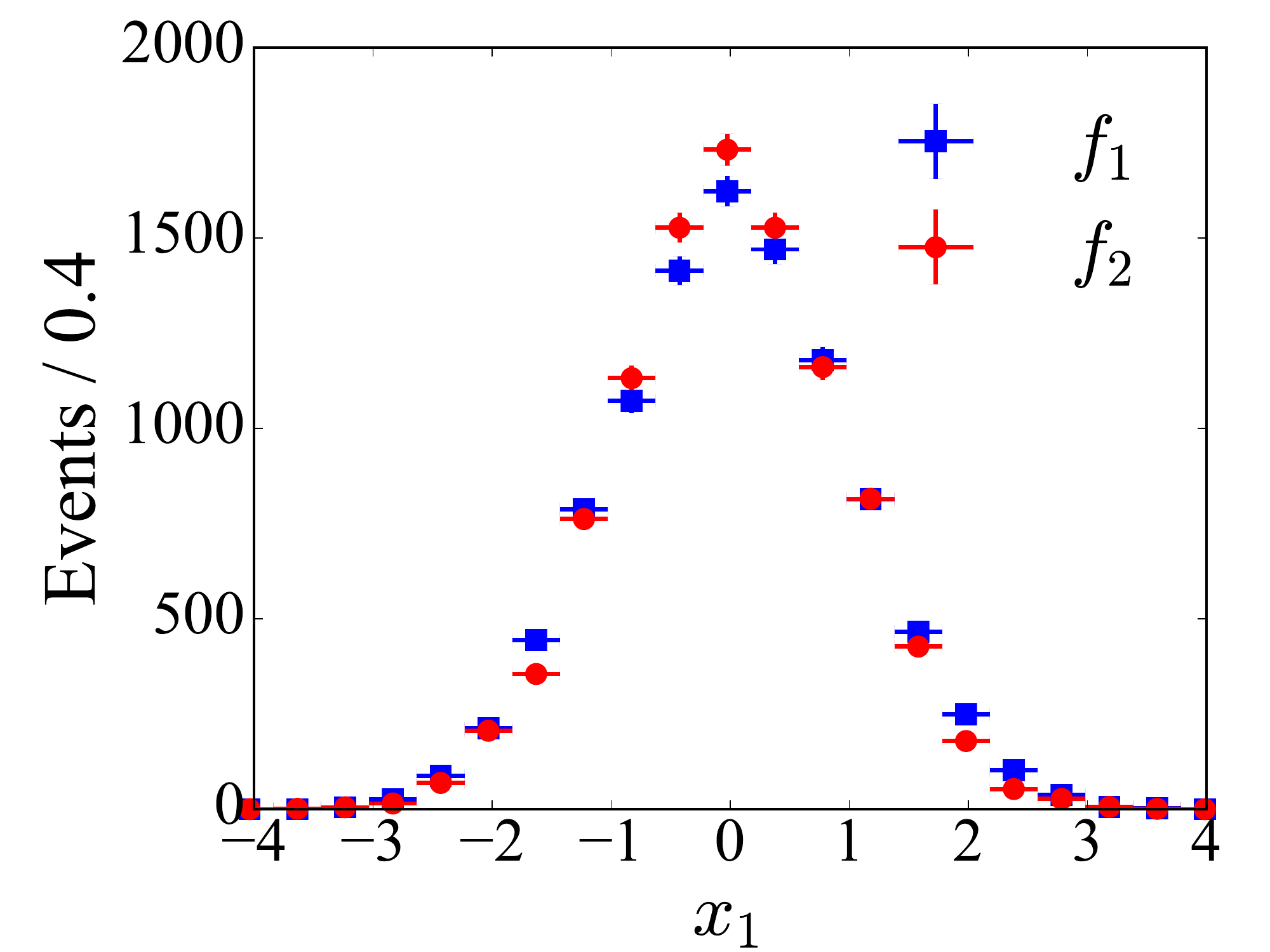}
  \includegraphics[width=0.49\textwidth]{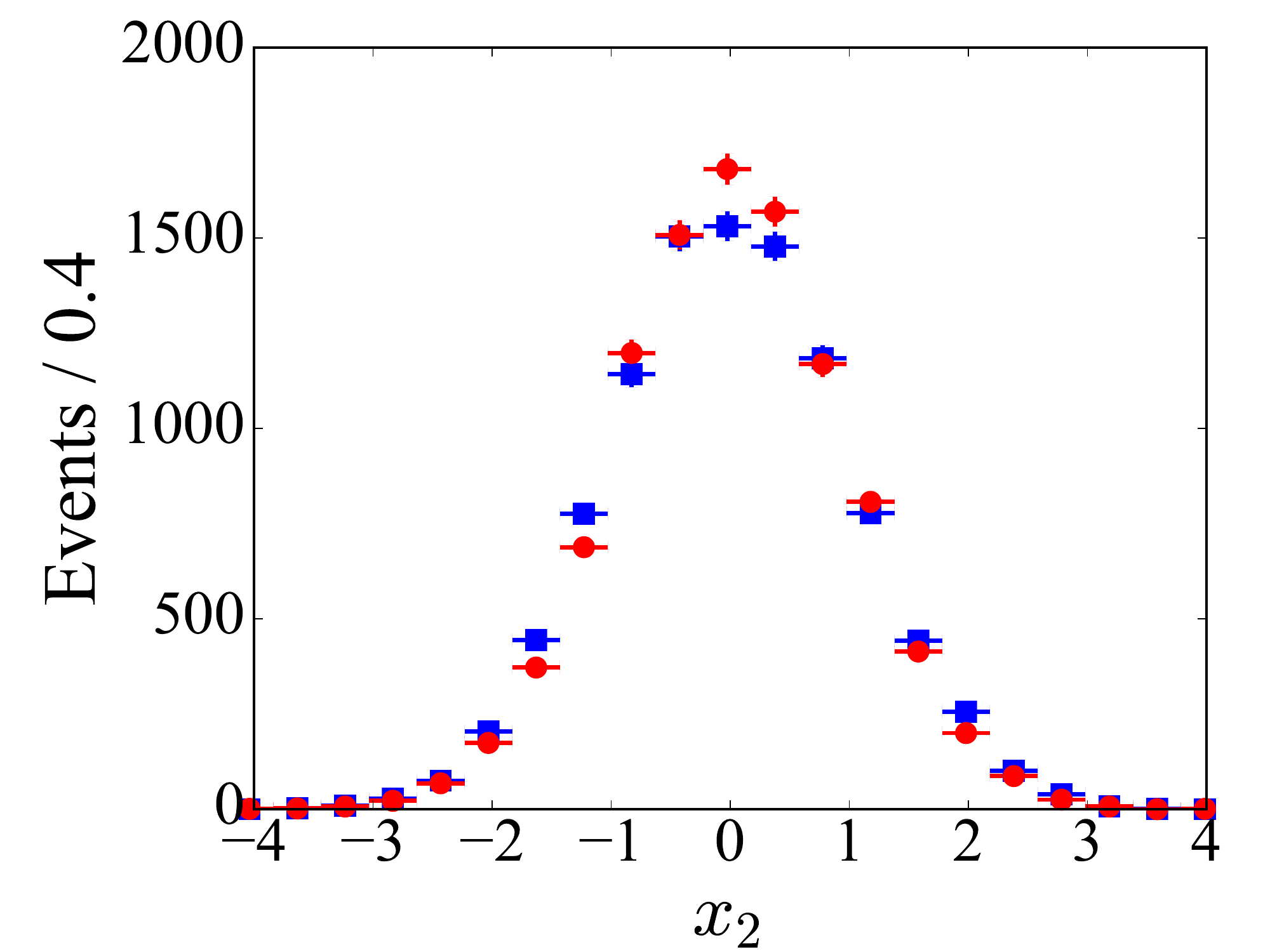}
  \caption{Gaussian example problem (shown for $d=2$): 
  (top left) a data set sampled from $f_1$;
  (top right) a data set sampled from $f_2$;
  and (bottom)
  projections of each data set onto the  $\hat{x}_1$ and $\hat{x}_2$ axes.}
  \label{fig:gauspdf}
\end{figure}

\begin{figure}
  \centering 
  \includegraphics[width=0.49\textwidth]{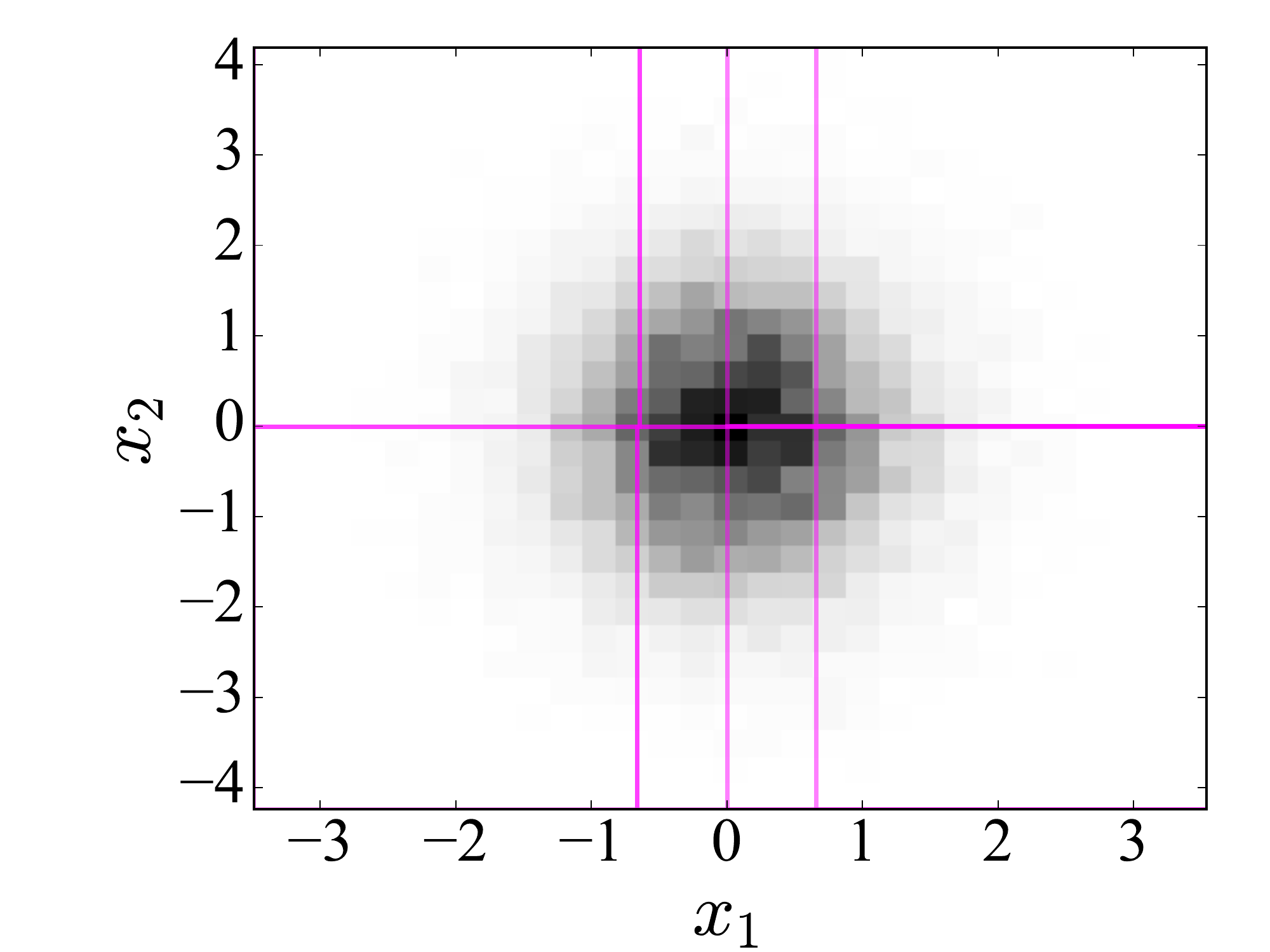}
  \caption{Gaussian example problem: Example optimal adaptive binning for a pooled $d=2$ data set.}
  \label{fig:gauschi2bins}
\end{figure}

\begin{figure}
  \centering 
  \includegraphics[width=0.32\textwidth]{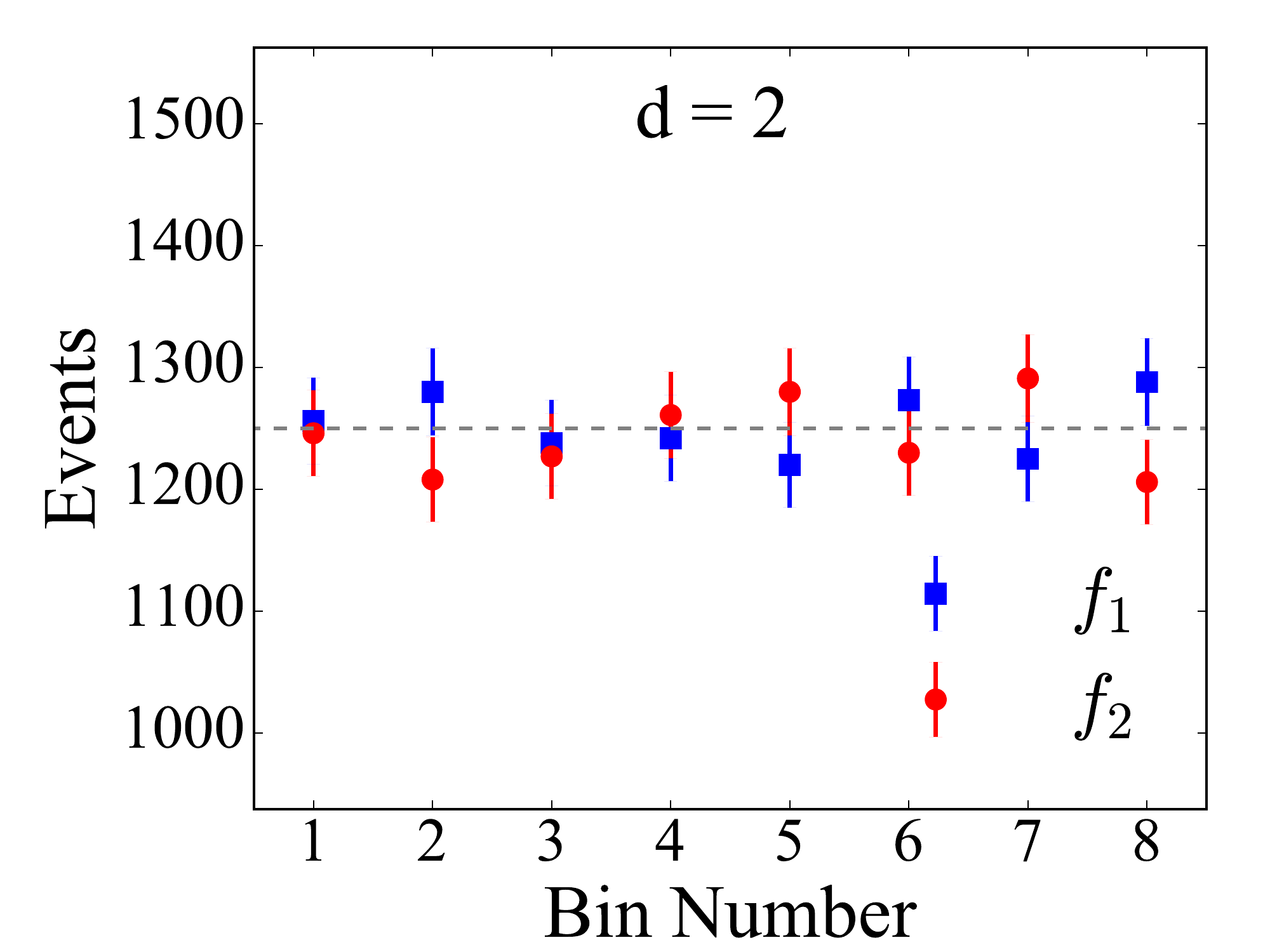}
  \includegraphics[width=0.32\textwidth]{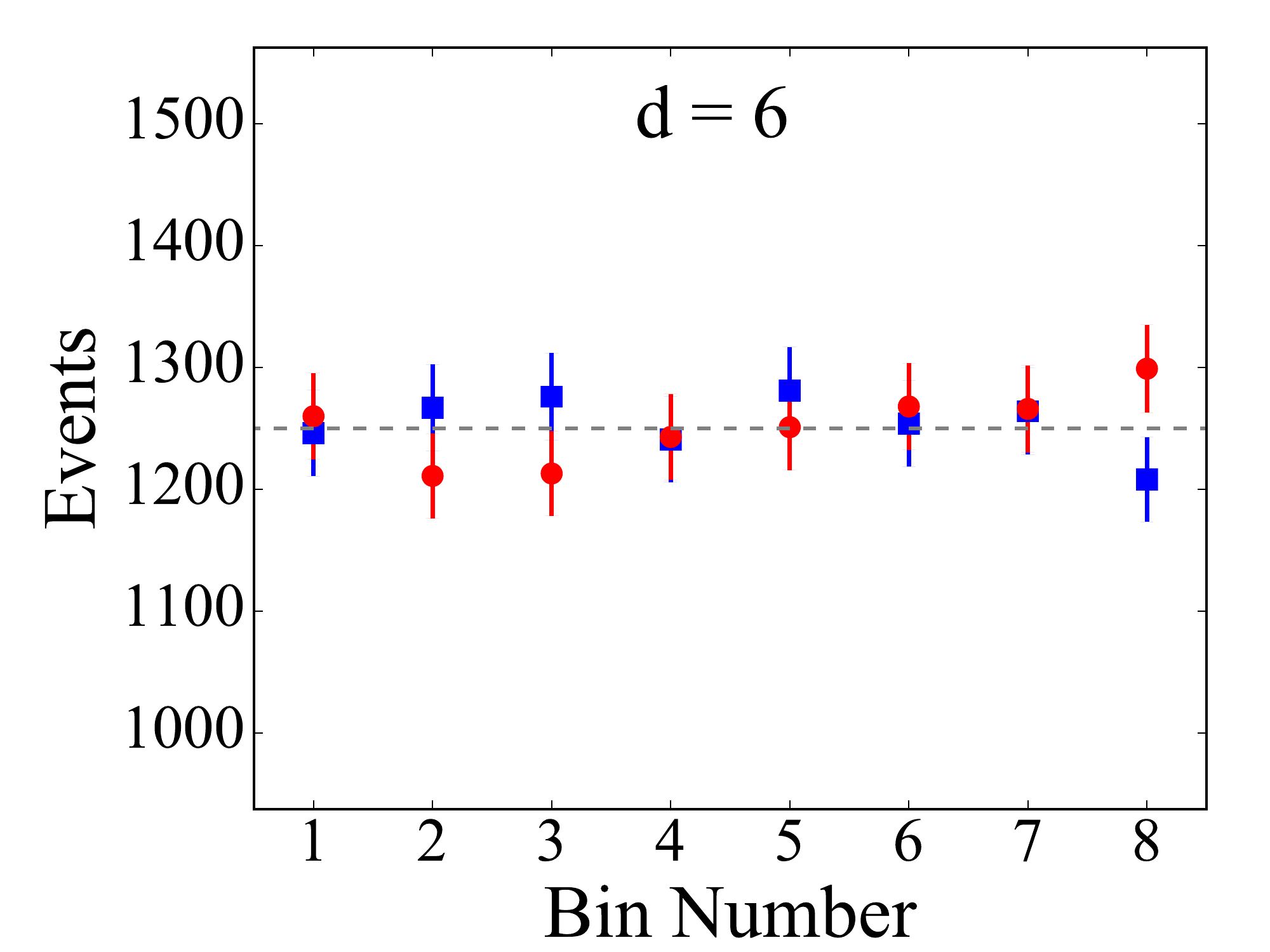}
  \includegraphics[width=0.32\textwidth]{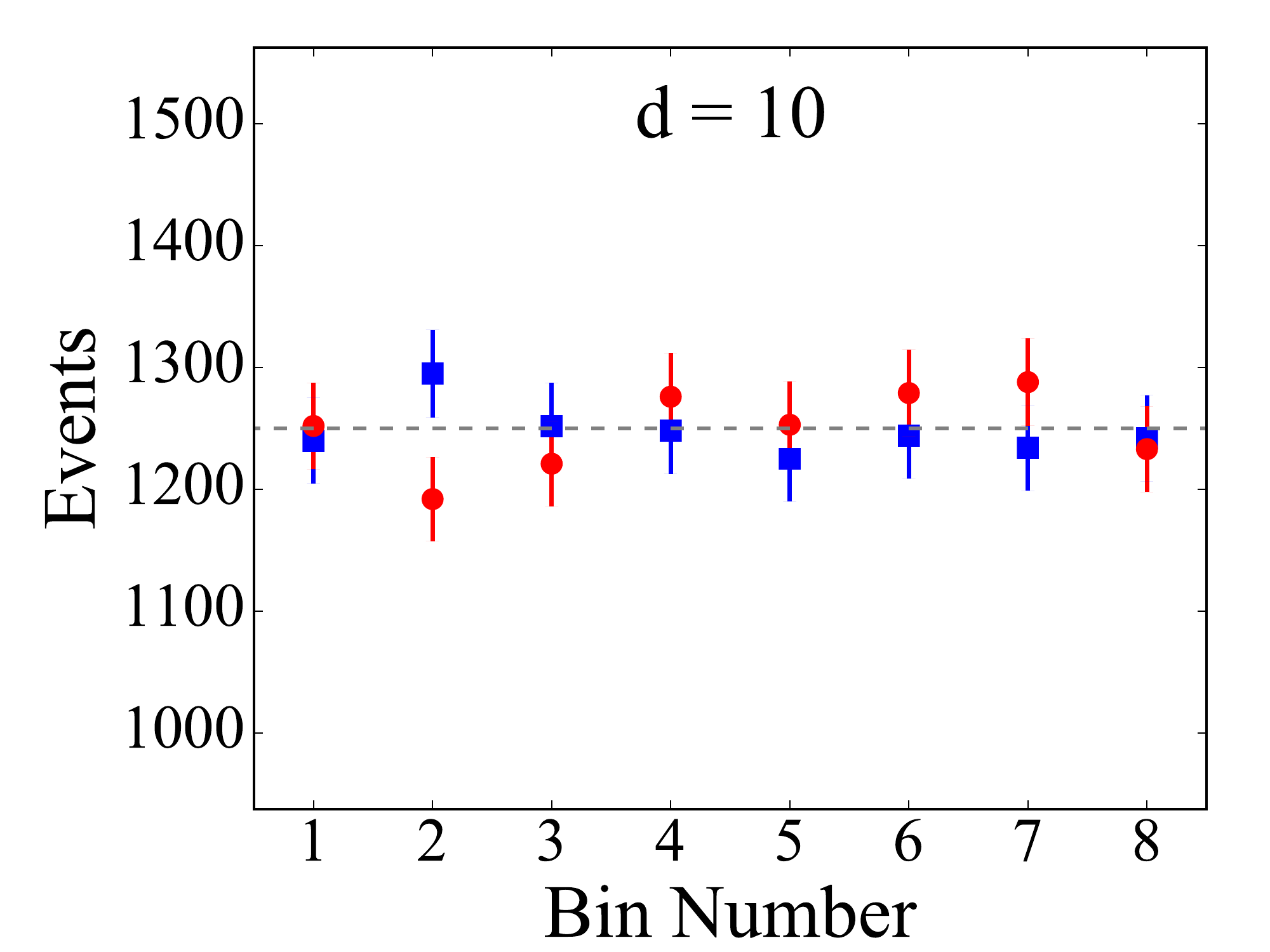}
  \caption{Gaussian example problem: adaptive-binned distributions (flattened into one dimension) for example data sets sampled from $f_1$ and $f_2$ for (left) $d=2$, (middle) $d=6$, and (right) $d=10$.}
  \label{fig:gauschi2}
\end{figure}

\begin{figure}
  \centering 
  \includegraphics[width=0.5\textwidth]{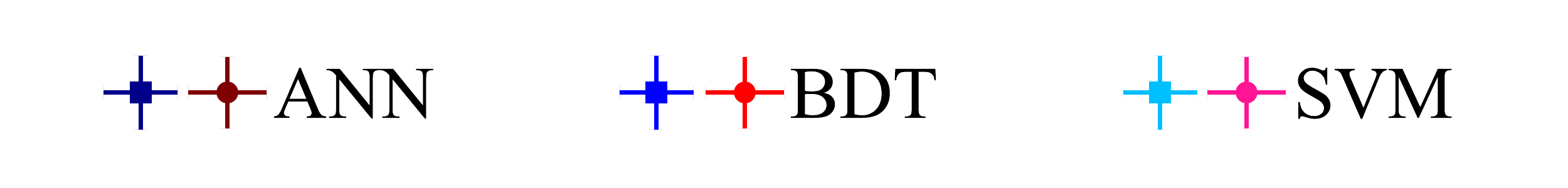} \vspace{-0.1in}\\
  \includegraphics[width=0.32\textwidth]{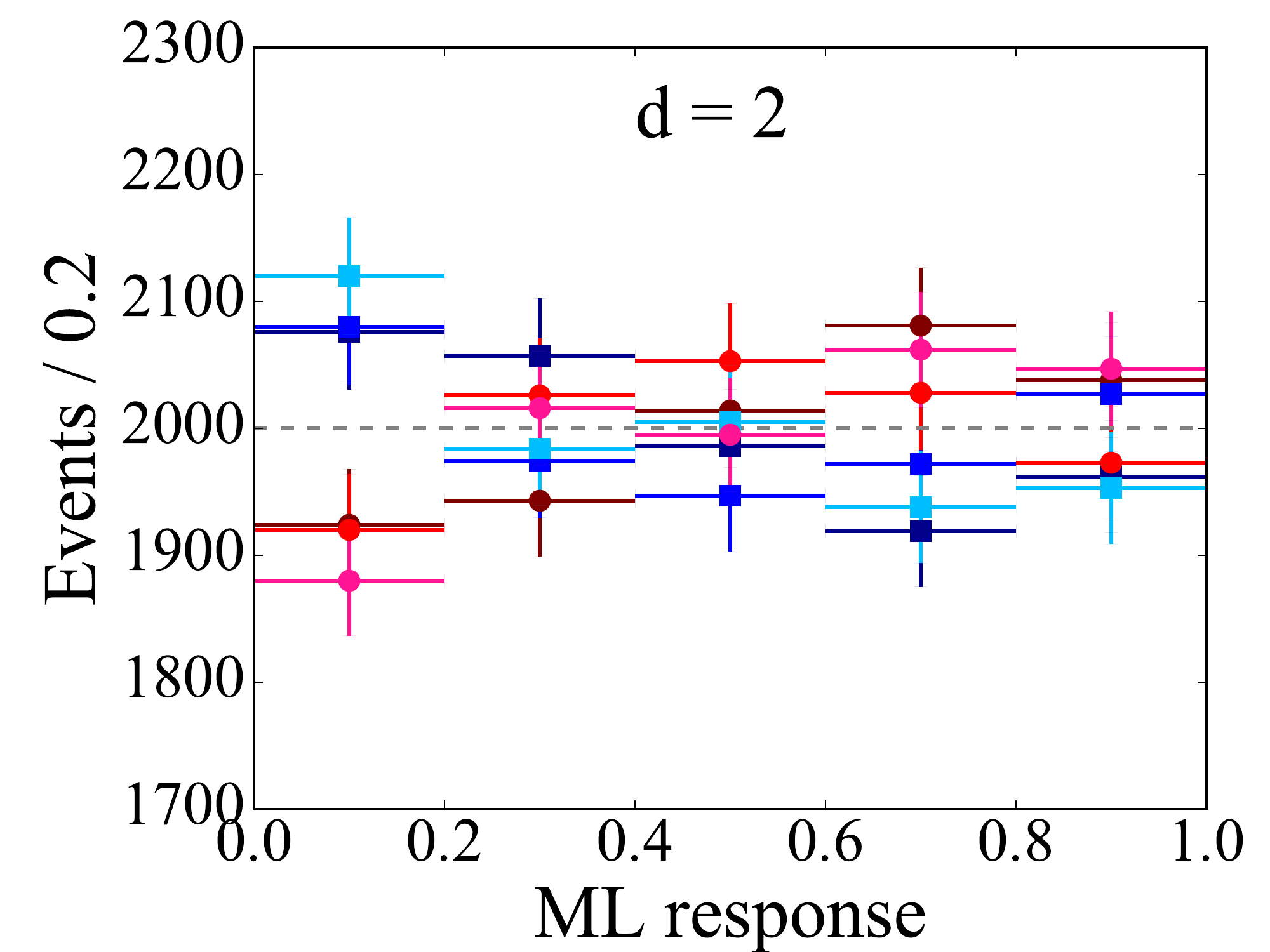}
  \includegraphics[width=0.32\textwidth]{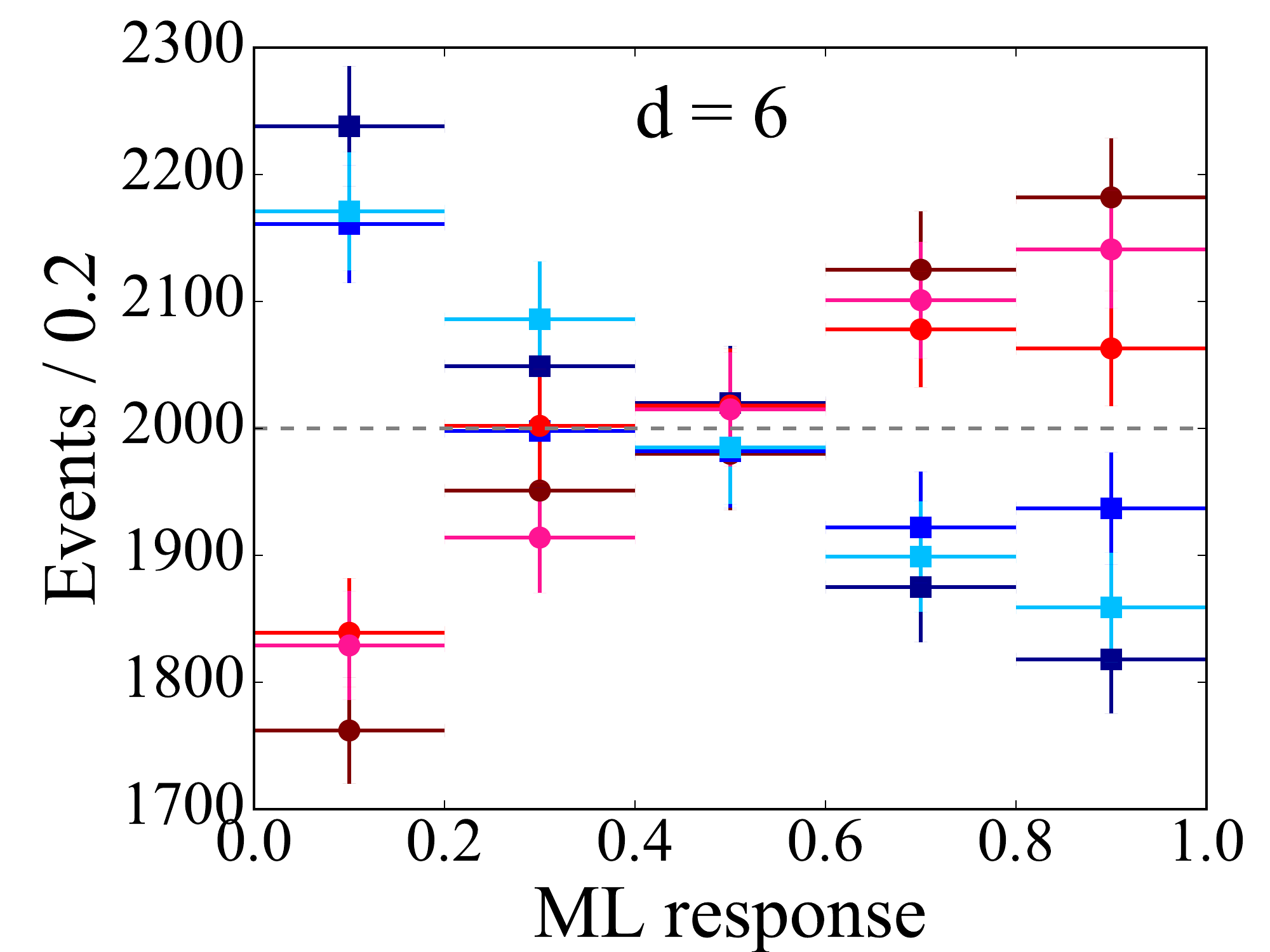}
  \includegraphics[width=0.32\textwidth]{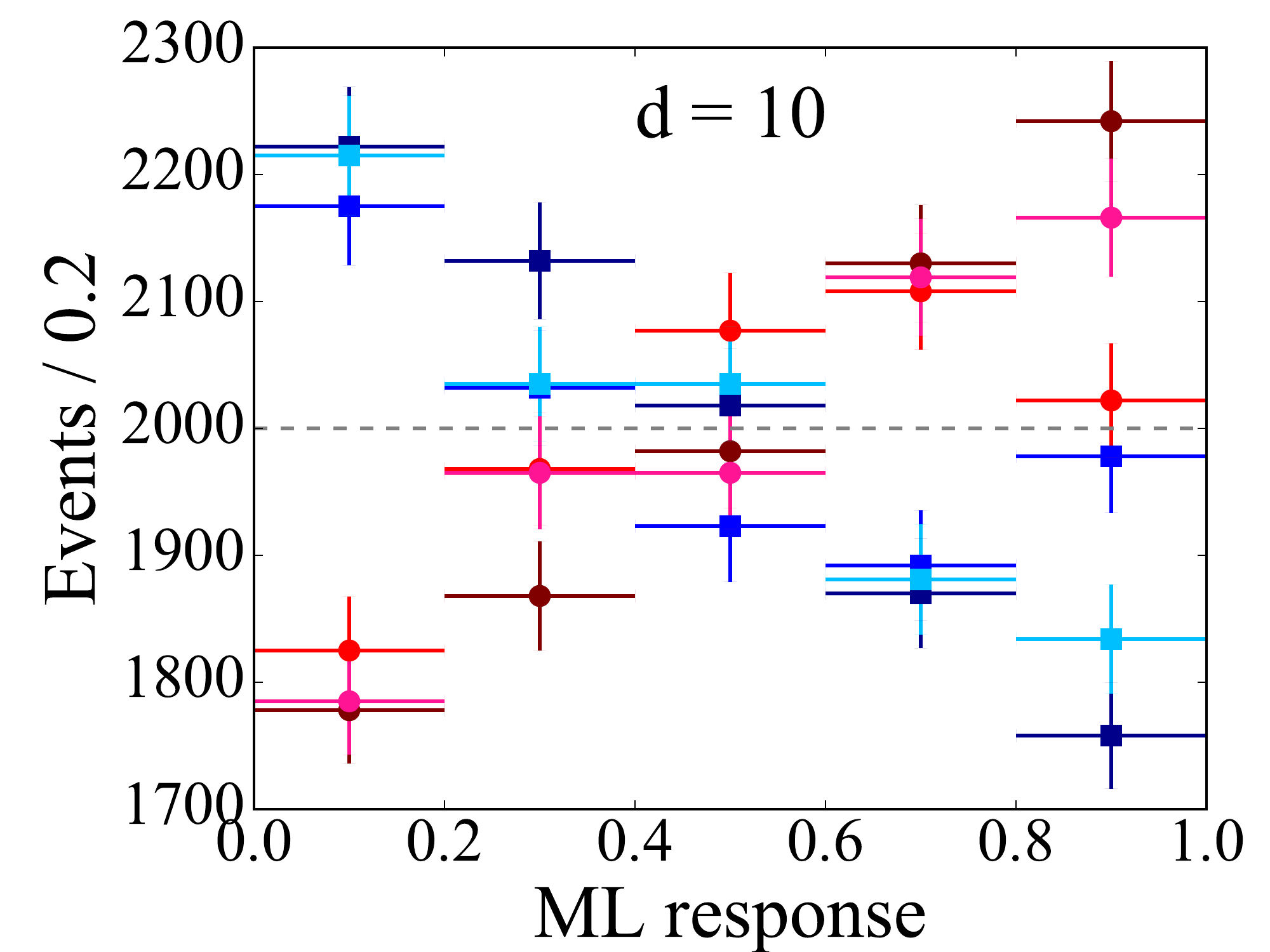}
  \caption{Gaussian example problem:  ML response distributions for example data sets sampled from $f_1$ and $f_2$ for (left) $d=2$, (middle) $d=6$, and (right) $d=10$.}
  \label{fig:gausml}
\end{figure}

First we consider a $d$-dimensional Gaussian distribution, where the values $\{x_k\}_{k=1}^{d}$ are the observed features for each event. 
Events are sampled from the $d$-dimensional PDF 
\begin{equation}
 f(\vx) = \prod\limits_{k=1}^{d} \mathcal{G}(x_k;\mu_k=0,\sigma_k),
\end{equation}
where $\sigma_k = 1$ for $f_1$ and $\sigma_k = 0.95$  for $f_2$. 
Figure~\ref{fig:gauspdf} shows example data sets sampled from $f_1$ and $f_2$ for the case $d=2$, as this is the only dimensionality that is easily displayed. 
Figure~\ref{fig:gauschi2bins} shows the optimal adaptive binning for an example two-dimensional data set, while Fig.~\ref{fig:gauschi2} shows example optimal-binned distributions for $d=2$, 6, and 10; the binned spectra are shown flattened into a single dimension by plotting versus bin number to aid in visually comparing the $a$ and $b$ samples. 
Example ML-response distributions are shown in Fig.~\ref{fig:gausml}. 

\begin{figure}
  \centering 
  \includegraphics[width=0.49\textwidth]{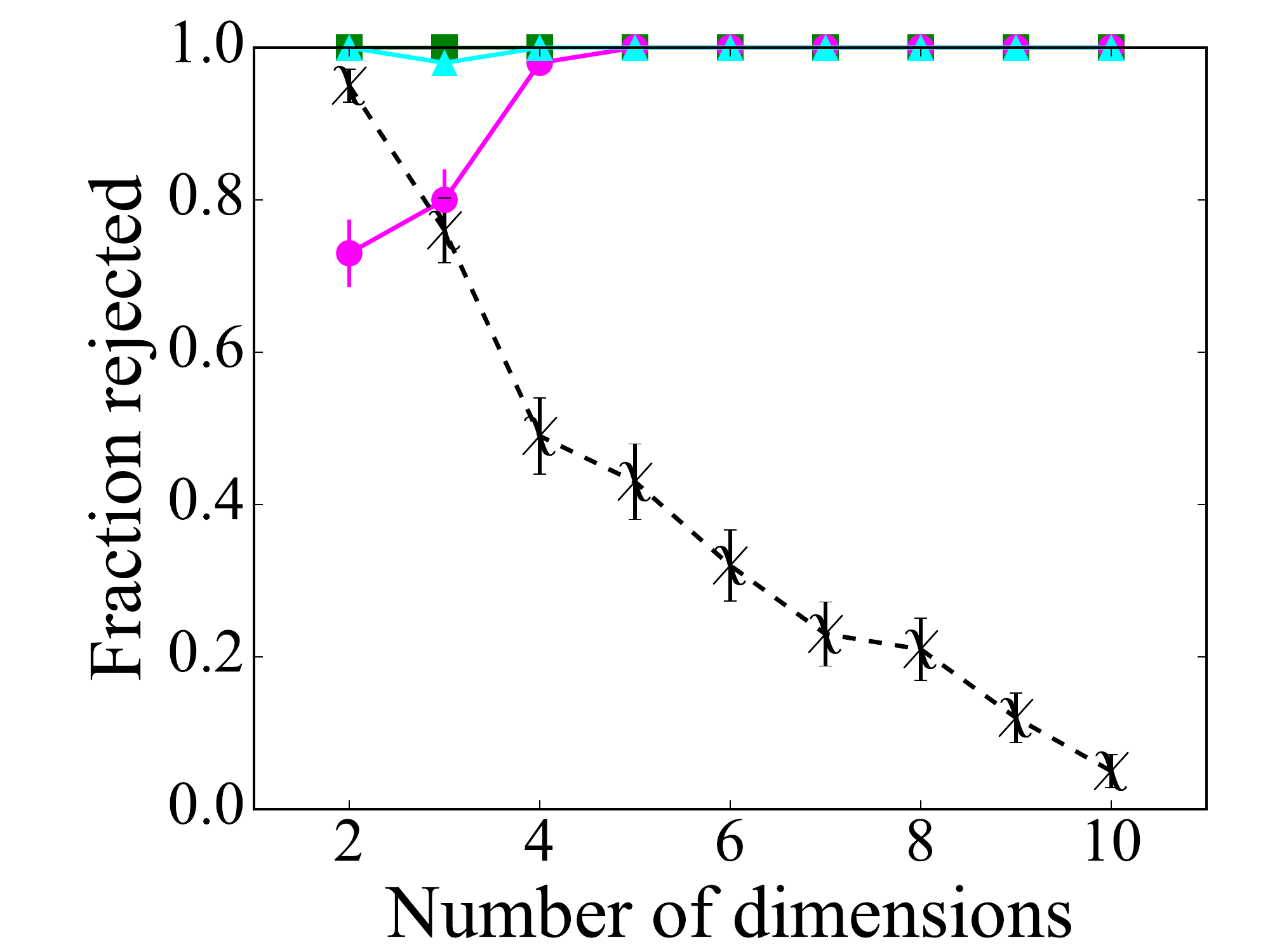}
  \includegraphics[width=0.49\textwidth]{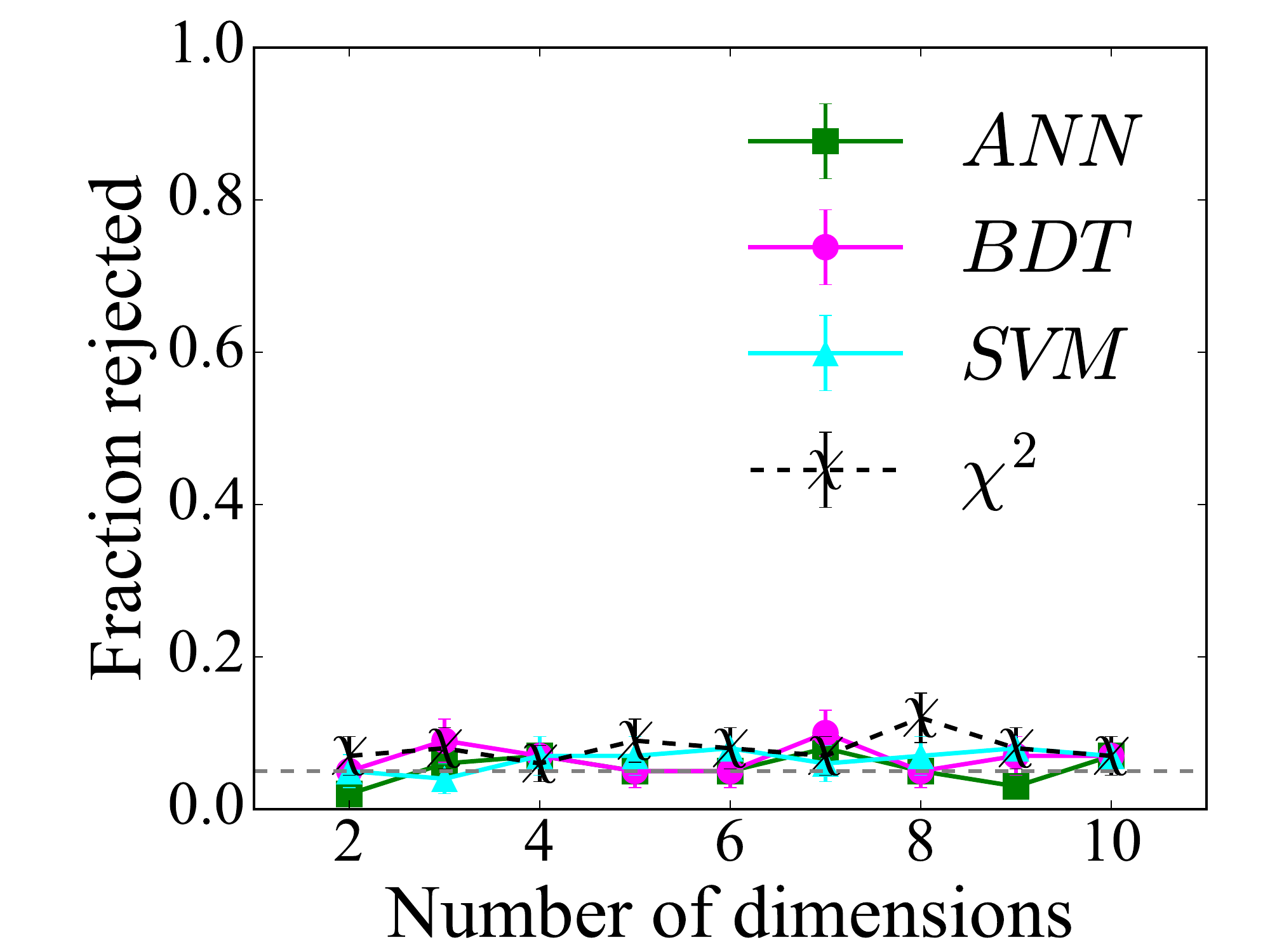}
  \caption{Gaussian example problem: Power of each method versus dimensionality $d$.}
  \label{fig:gaus}
\end{figure}

First, we verify that each test produces valid $p$-values under the null hypothesis by generating an ensemble of $2\times100$ data sets where $f_a = f_b = f_1$. 
Figure~\ref{fig:gaus} shows that both the multivariate $\chi^2$ and ML-based methods reject the expected $\approx 5\%$ of data samples when the null hypothesis is true. 
Next, we generate an alternative set of 100 data sets with $f_b = f_2$. 
Figure~\ref{fig:gaus} shows that the ML-based tests outperforms the $\chi^2$ test for $d > 2$.
In this example, where the PDFs are hyperspherical, the ANN and SVM outperform the BDT.
This is not surprising, since decision trees are constructed through a series of one-dimensional (rectangular) splittings of the data.  
The main conclusion drawn from this example problem is that, as expected,
the power of the $\chi^2$ test decreases rapidly as $d$ increases, while the power of the ML-based tests is much less sensitive to dimensionality.

\begin{figure}[t]
  \centering 
  \includegraphics[width=0.49\textwidth]{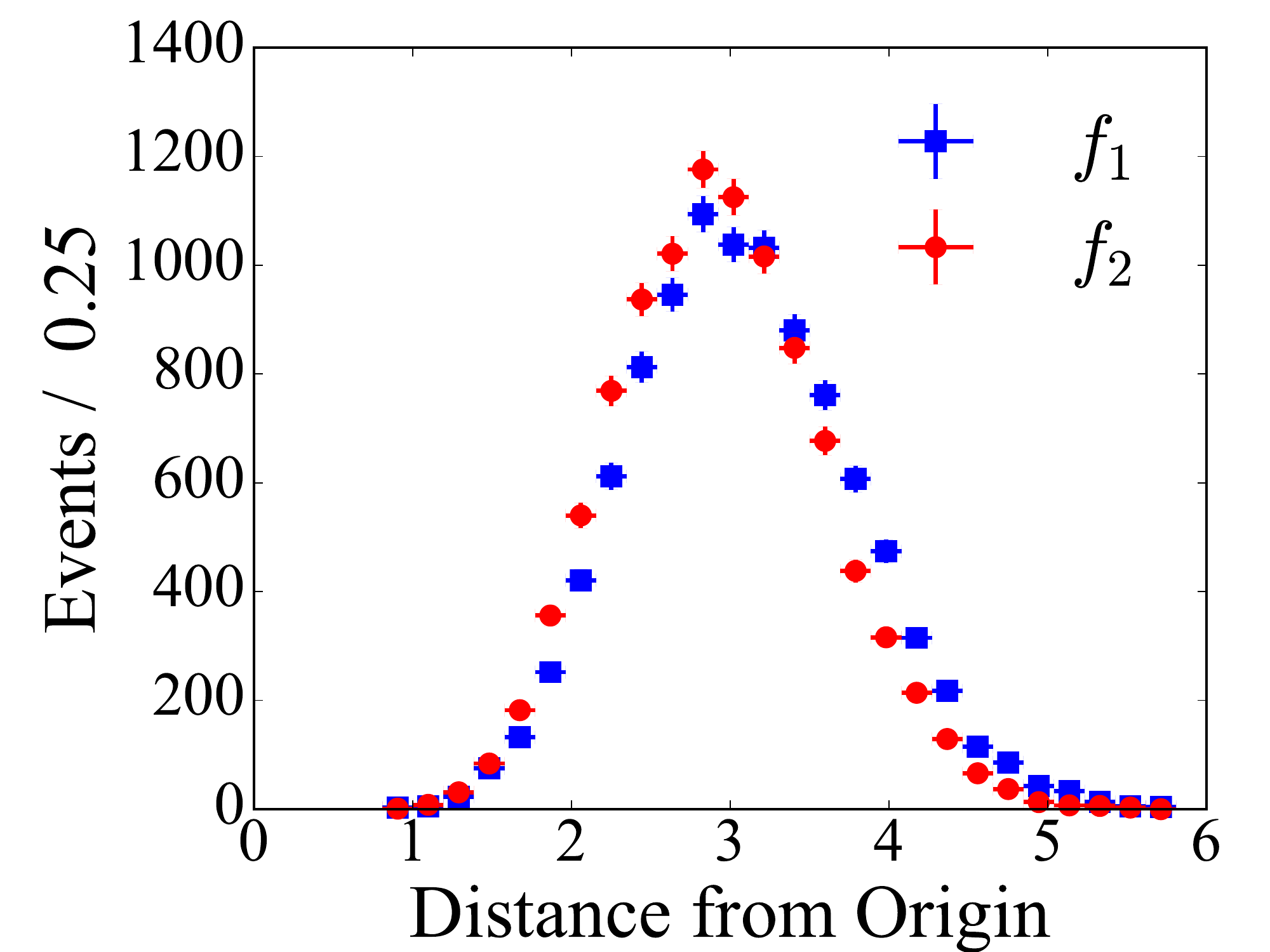}
  \includegraphics[width=0.49\textwidth]{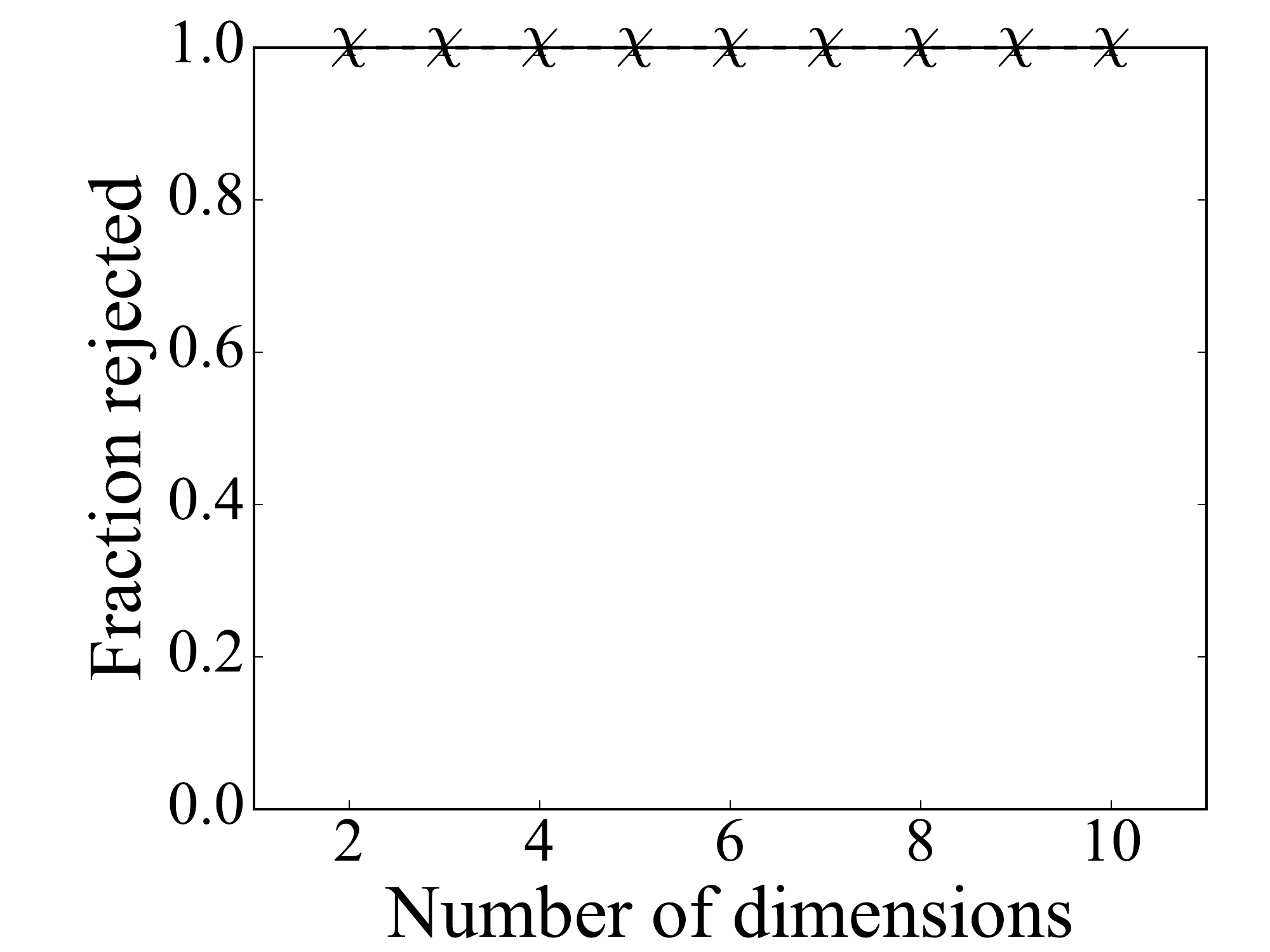}
  \caption{Gaussian example problem reduced to 1-D using $d$-dimensional Euclidean distance: 
  (left) Euclidean distance for a 10-dimensional data set;
  and 
  (right) the power of applying a one-dimensional $\chi^2$ test on the reduced problem versus $d$. }
  \label{fig:gaus_euc}
\end{figure}

If the underlying model is known, we can design a powerful feature for distinguishing between $f_1$ and $f_2$.  Figure~\ref{fig:gaus_euc} shows the $d$-dimensional Euclidean distance from the origin of each measurement for $d=10$.  This feature optimally reduces this $d$-dimensional problem into a one-dimensional one.  
Applying the one-dimensional $\chi^2$ test to the reduced problem provides greater power than applying the ML-based approach to the $d$-dimensional one.  
Indeed, this is a common approach in physics analyses, {\em e.g.}, when studying particle decays the invariant mass of the decay products is used to reduce all of the kinematic information  into a single feature. \footnote{The popularity of this approach does not mean that information is not lost when using it, {\em e.g.}, see Ref.~\cite{Baldi:2014pta}.}
Our $d$-dimensional Gaussian example problem is effectively a one-dimensional one, and so the adaptive-binned $\chi^2$ is a powerful option if we treat it as such; however, if we are unable to design a feature to fully capture all of the relevant information, which is likely the case when the underlying model is unknown, then the ML-based approach is more powerful than traditional ones.

\subsection{Oscillatory}

\begin{figure}[t]
  \centering 
  \includegraphics[width=0.49\textwidth]{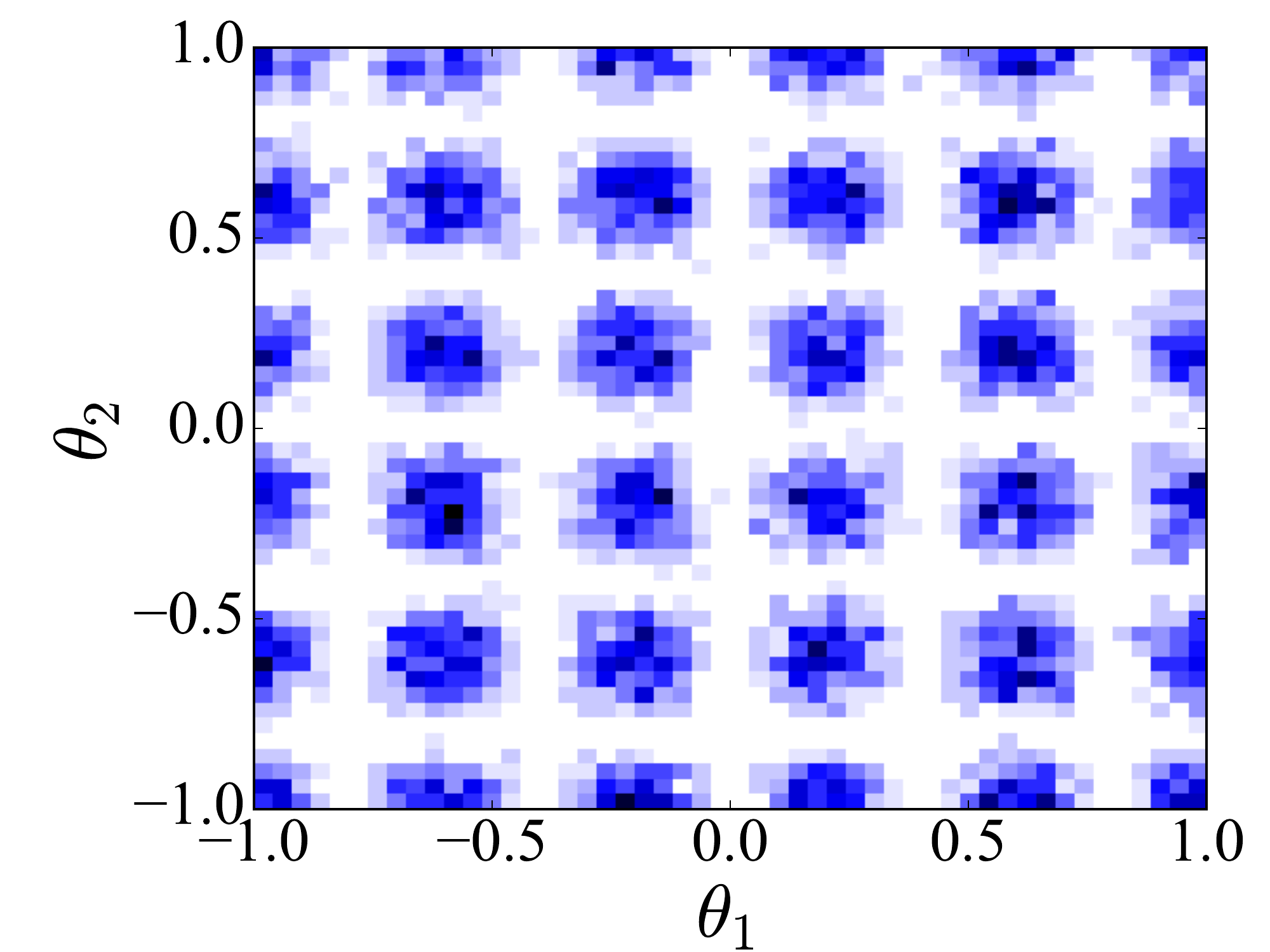}
  \includegraphics[width=0.49\textwidth]{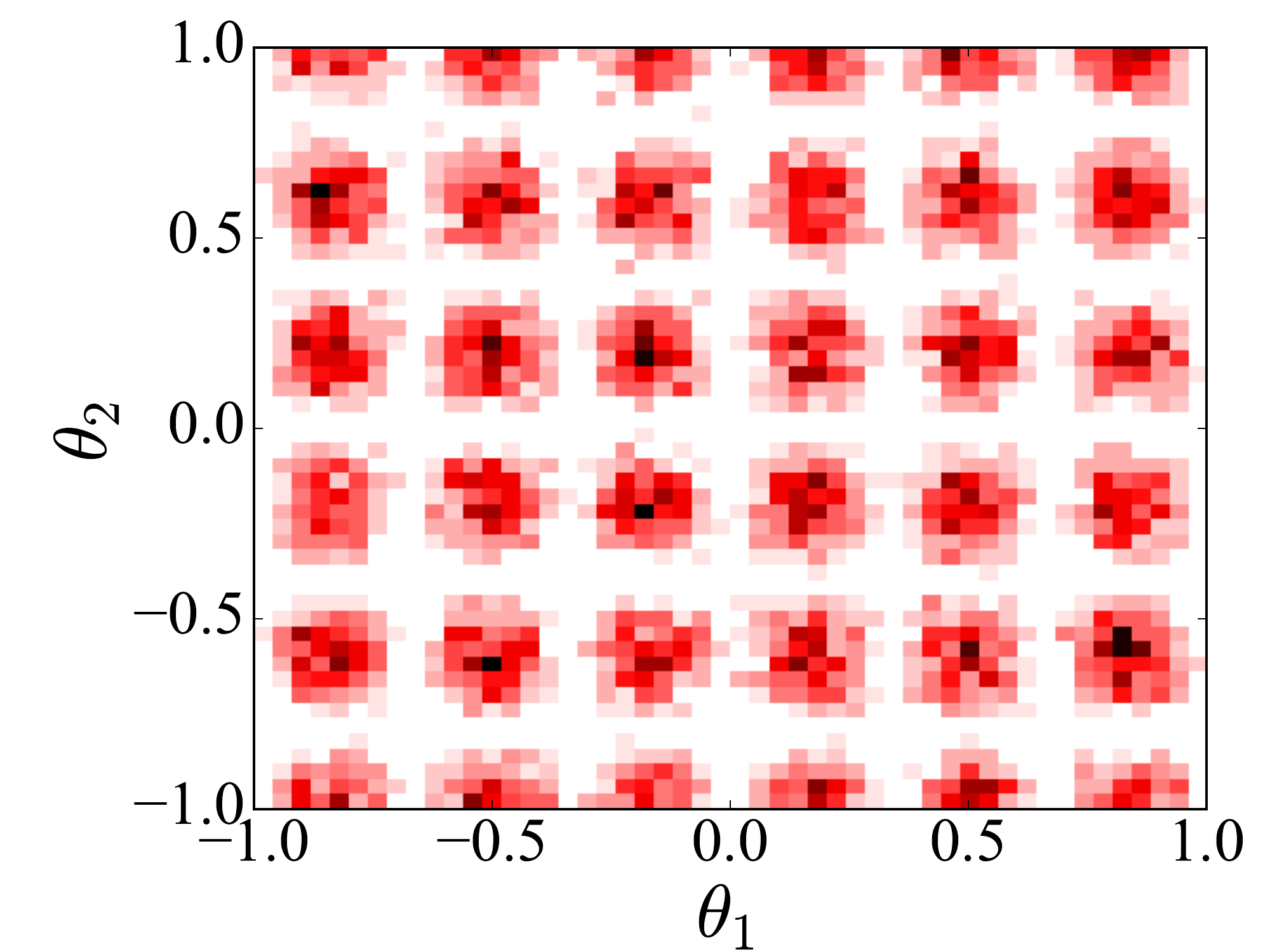}
  \caption{Oscillatory example problem (shown for $d=2$): 
  (left) a data set sampled from $f_1$ and 
  (right) a data set sampled from $f_2$.}
  \label{fig:oscpdf}
\end{figure}

\begin{figure}[t]
  \centering 
  \includegraphics[width=0.49\textwidth]{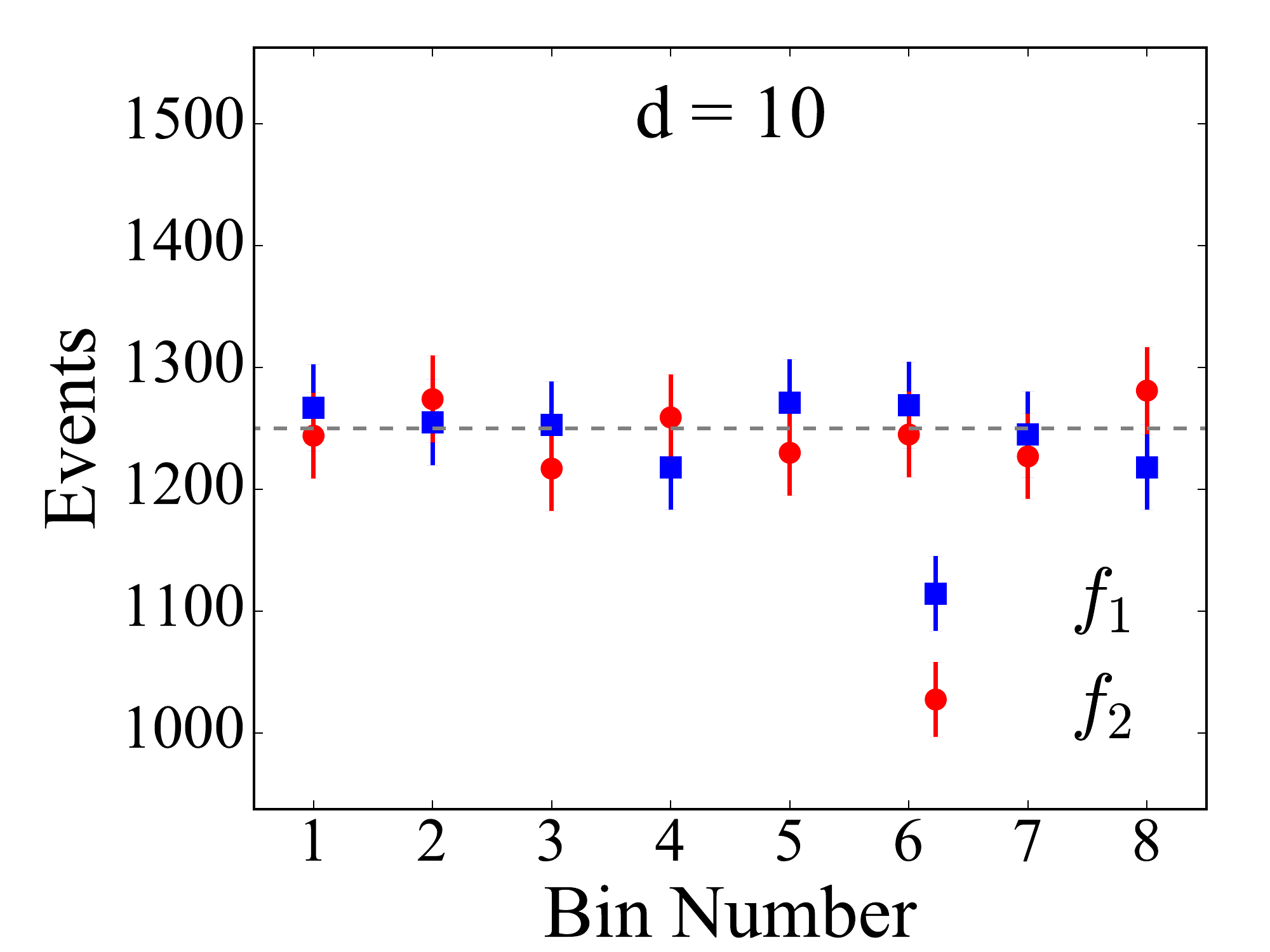}
 \includegraphics[width=0.49\textwidth]{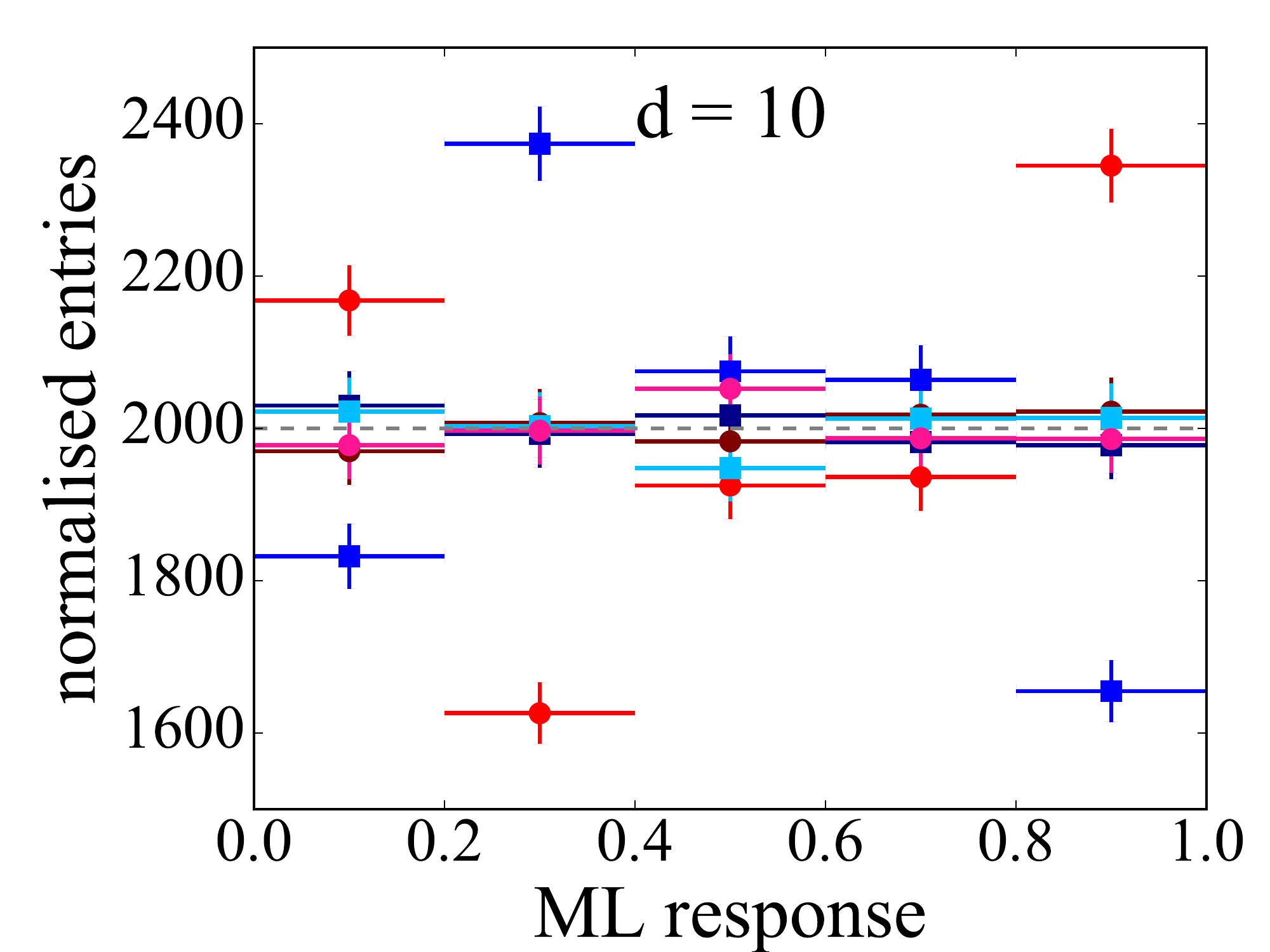}
  \caption{Oscillatory example problem:  (left) $\chi^2$ and (right) ML response distributions for example data sets sampled from $f_1$ and $f_2$ for $d=10$ (same marker scheme as in Fig.~4).}
  \label{fig:oscml}
\end{figure}

\begin{figure}
  \centering 
  \includegraphics[width=0.49\textwidth]{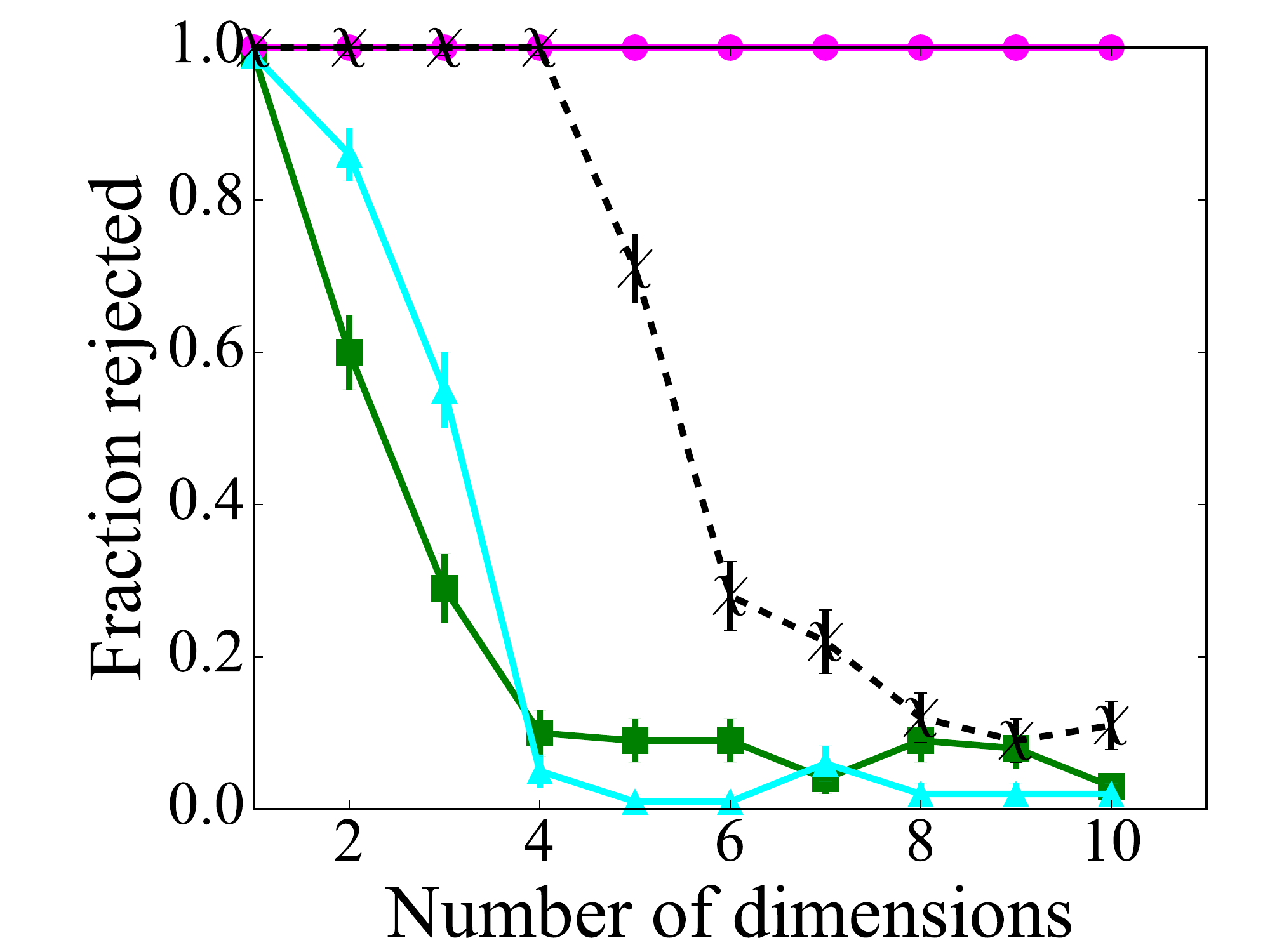}
   \includegraphics[width=0.49\textwidth]{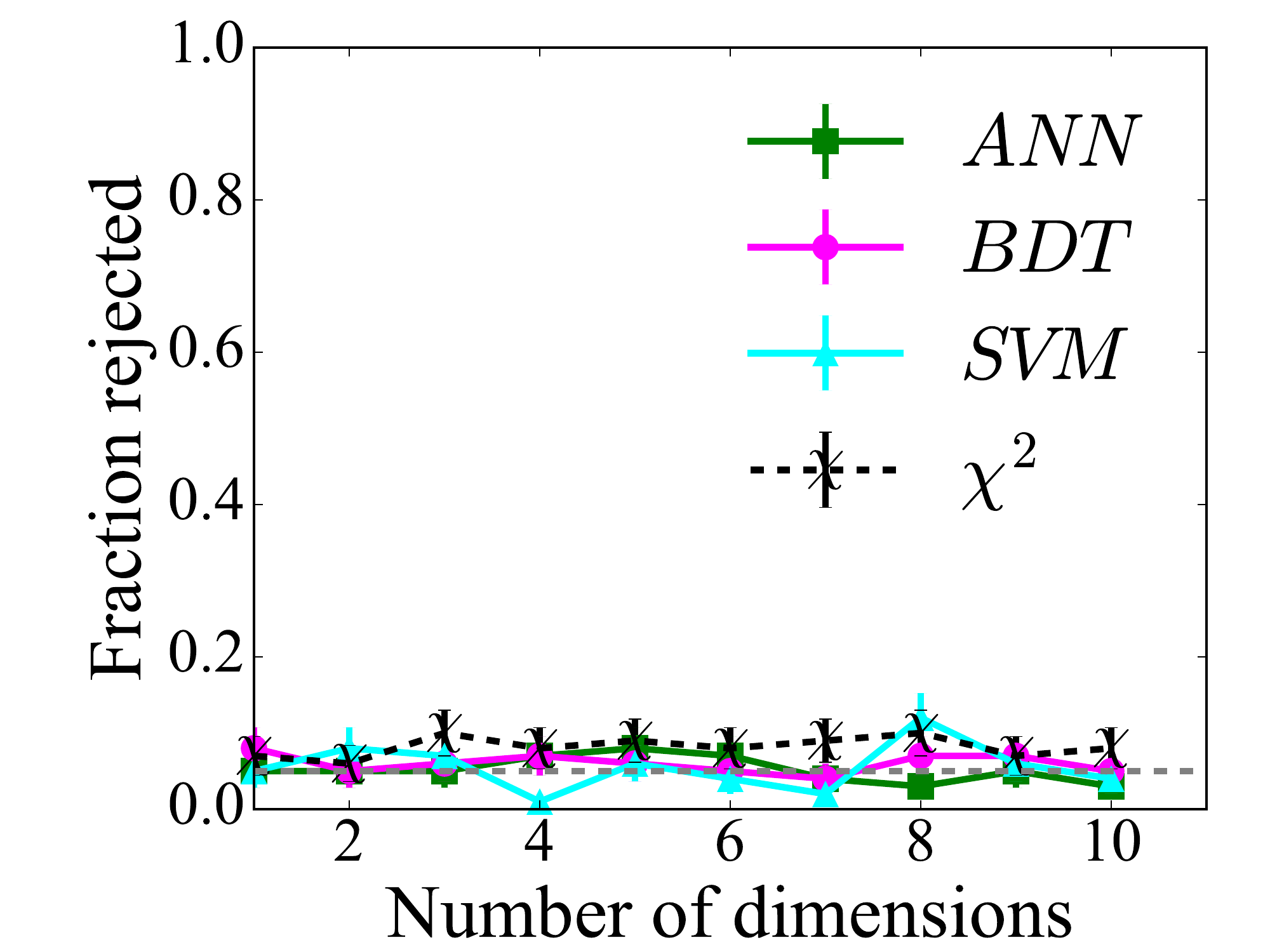}
  \caption{Oscillatory example problem:  Power of each method versus dimensionality $d$.}
  \label{fig:osc}
\end{figure}

Next we consider a $d$-dimensional oscillatory problem where the observed features are $\vx = \vec{\theta} = \{\theta_k\}_{k=1}^d$ and the PDF is defined according to
\begin{equation}
f(\vec{\theta}) = \prod\limits_{k=1}^{d} \sin^2{\left(n_k \theta_k\right)},
\end{equation}
where $n_k = 5$ for all $k$ in $f_1$, but for $f_2$ the value of $n_1 = 6$ is used (all other $n_{k>1} =5$, see Fig.~\ref{fig:oscpdf}).  
The results of applying $\chi^2$ and ML-based two-sample tests are shown in Figs.~\ref{fig:oscml} and \ref{fig:osc}. 
Here, the underlying rectangular nature of the BDT is advantageous, whereas our chosen set of ANN and SVM hyperparameters do not lead to good performance of these algorithms on this problem.  
This example demonstrates that in some cases, one ML algorithm may vastly outperform some others. 
Studying the performance of different ML-based tests on simulated data samples can reveal such situations. 
We note that this problem can also be reduced to a single dimension by projecting onto the $\theta_1$ axis, if one knows that the two models are the same in all other dimensions. 
The fact that our first two example problems possess this feature should not be taken to mean that most real-world problems are inherently one-dimensional. Rather, the simplest problems to implement for any dimensionality $d$ are inherently one-dimensional.

\subsection{Particle Decays}

\begin{figure}[t]
         \centering 
  \includegraphics[width=0.49\textwidth]{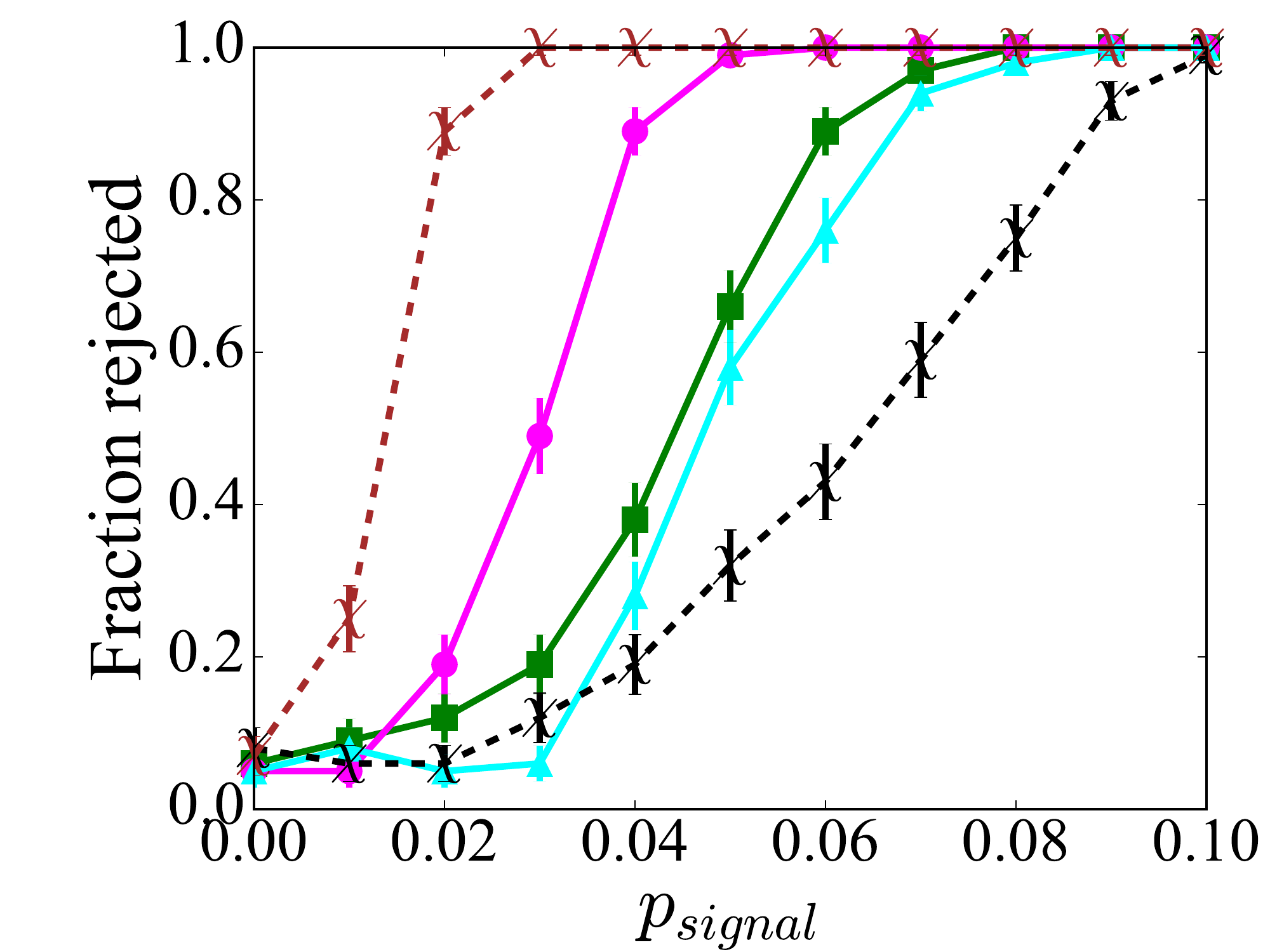}
  \caption{Particle-decay problem:  Power of each method versus relative signal component size.  The marker scheme is the same as Figs.~5 and 9, with an additional red $\chi$ marker added that denotes performing the $\chi^2$ test on the one-dimensional invariant-mass distribution.}
  \label{fig:mass}
\end{figure}

For the final example problem, we consider a simple real-world physics use case of searching for a particle of mass $m(X)$ decaying into two equal-mass lighter particles $X \to \alpha \beta$. 
The energy and momentum of the particles $\alpha$ and $\beta$ are measured in a detector producing an 8-dimensional feature space, where the resolution on the invariant mass of the $\alpha\beta$ final state, $m(\alpha\beta)$,  is $0.1\,m(X)$. 
There is a background present that is uniformly distributed in $m(\alpha\beta)$ on $[0.5\,m(X),1.5\,m(X)]$.
We have a background-only data sample (in a real search, this could be from simulation, same-sign candidates in data, {\em etc.}) and another data sample in which we want to look for evidence of an $X \to \alpha \beta$ signal.
First, we apply the ML-based and adaptive-binned $\chi^2$ tests to the 8-dimensional problem, where the fraction of the search data sample that arises from $X \to \alpha \beta$ decays is varied from zero to 10\%.
The results are shown in Fig.~\ref{fig:mass}. 
As expected, the ML-based methods outperform the $\chi^2$ test; however, when studying particle decays, we know that calculating the invariant mass is an effective way to reduce the kinematic information into a single feature. 
Figure~\ref{fig:mass} also shows that, for this simple problem, the one-dimensional $\chi^2$ test on the $m(\alpha\beta)$ distribution outperforms the ML-based methods performed on the  8-dimensional $\{E_{\alpha},\vec{p}_{\alpha},E_{\beta},\vec{p}_{\beta}\}$ feature space --- at least for the ML algorithms employed in this study. 
More advanced deep-learning algorithms may be competitive with the human-led feature design approach of using the invariant mass here~\cite{Baldi:2014pta}.

\section{Summary}
\label{sec:sum}

Multivariate goodness-of-fit and two-sample tests are important components of many nuclear and particle physics analyses; however, the tests commonly used in such analyses are only powerful when the dimensionality is small. 
Machine learning classifiers are powerful tools capable of reducing highly multivariate problems into univariate ones, on which commonly used tests such as $\chi^2$ or Kolmogorov-Smirnov may be applied. 
We explored applying both traditional and machine-learning-based tests to several example problems, and studied how the power depends on dimensionality. 
For up to about three dimensions, the power of the adaptive-binned $\chi^2$ test is comparable to that of ML-based tests; however, for higher dimensionality the ML-based approach is far superior to the $\chi^2$ test.

A caveat to the previous statement is that if some information about the underlying model is known,  it may be possible to construct features that reduce the effective dimensionality of the problem.  
Indeed, such human-led feature design is common in physics analyses, {\em e.g.}, when studying particle decays the invariant mass is used to reduce all of the kinematic information into a single feature. 
In such cases, the problem may effectively be a low-dimensional one and the adaptive-binned $\chi^2$ is a powerful option.
When one or two features cannot be designed to fully capture all of the relevant information --- which is likely the case when detector-response features are included in the problem ---  ML-based tests are expected to be more powerful than traditional ones.
We encourage the community to investigate adopting this approach to goodness-of-fit and two-sample testing for problems where the dimensionality is large and cannot be reduced without significant loss of information.  

\acknowledgments

We thank Kyle Cranmer for helpful feedback. 
This work was supported by DOE grants DE-SC0010497 and DE-FG02-94ER40818.

\bibliographystyle{h-physrev}
\bibliography{refs}

\end{document}